\documentclass[12pt]{article}%,letterpaper,notitlepage,dash,spanish]{article}
\usepackage{amsmath}
\usepackage{amsthm}
\usepackage{amsfonts}
\usepackage[dvips]{epsfig}
\usepackage{graphicx}
\usepackage{amssymb}
\usepackage{cancel}
\usepackage{pstricks} 
\usepackage{lscape}
\usepackage{enumitem}
\usepackage{subcaption}

\usepackage[sort&compress,numbers, merge]{natbib}

\DeclareFontFamily{U}{mathx}{\hyphenchar\font45}
\DeclareFontShape{U}{mathx}{m}{n}{<-> mathx10}{}
\DeclareSymbolFont{mathx}{U}{mathx}{m}{n}
\newcommand{\beq}{\begin{equation}}
\newcommand{\eeq}{\end{equation}}

\makeatletter
\newlength{\apb@width}
\newcommand{\autoparbox}[2][c]{\settowidth{\apb@width}{#2}\parbox[#1]{\apb@width}{#2}}

\newcommand{\Cen}[2]{%
  \ifmeasuring@
    #2%
  \else
    \makebox[\ifcase\expandafter #1\maxcolumn@widths\fi]{$\displaystyle#2$}%
  \fi
}
\makeatother

\definecolor{Orange}{cmyk}{0,0.61,0.87,0}
\definecolor{JungleGreen}{cmyk}{0.99,0,0.52,0}
\definecolor{OliveGreen}{cmyk}{0.64,0,0.95,0.40}
\definecolor{Brown}{cmyk}{0,0.81,1,0.60}
\definecolor{RoyalBlue}{cmyk}{0.71,0.53,0,0.12}

\allowdisplaybreaks[1]

\newcommand{\1}{\mbox{1}\hspace{-0.25em}\mbox{l}}

\renewcommand{\arraystretch}{1.3}

\usepackage[colorlinks=true, linkcolor=OliveGreen, citecolor=RoyalBlue,
urlcolor=RoyalBlue]{hyperref}

\begin{document}

\vspace{-0.2in}
\begin{flushright}
{\tt KCL-PH-TH/2019-80}, {\tt CERN-TH-2019-175}  \\
{\tt UT-19-25, ACT-06-19, MI-TH-1938} \\
{\tt UMN-TH-3901/19, FTPI-MINN-19/24} \\
{\tt IFT-UAM/CSIC-19-137}
\end{flushright}

%\vspace{0.05cm}
\begin{center}
{\bf {\large Superstring-Inspired Particle Cosmology:} \\
\vspace{0.1cm}
Inflation, Neutrino Masses, Leptogenesis, Dark Matter \& the SUSY Scale}
\end{center}
\vspace{0.05cm}

\begin{center}{%\large
{\bf John~Ellis}$^{a}$,
{\bf Marcos~A.~G.~Garcia}$^{b}$,
{\bf Natsumi Nagata}$^{c}$, \\[0.1cm]
{\bf Dimitri~V.~Nanopoulos}$^{d}$ and
{\bf Keith~A.~Olive}$^{e}$
}
\end{center}

\begin{center}
{\em $^a$Theoretical Particle Physics and Cosmology Group, Department of
  Physics, King's~College~London, London WC2R 2LS, United Kingdom;\\
Theoretical Physics Department, CERN, CH-1211 Geneva 23,
  Switzerland;\\
National Institute of Chemical Physics and Biophysics, R\"{a}vala 10, 10143 Tallinn, Estonia}\\[0.2cm]
  {\em $^b$Instituto de F\'isica Te\'orica (IFT) UAM-CSIC, Campus de Cantoblanco, 28049, Madrid, Spain}\\[0.2cm] 
  {\em $^c$Department of Physics, University of Tokyo, Bunkyo-ku, Tokyo
 113--0033, Japan}\\[0.2cm] 
{\em $^d$George P. and Cynthia W. Mitchell Institute for Fundamental
 Physics and Astronomy, Texas A\&M University, College Station, TX
 77843, USA;\\ 
 Astroparticle Physics Group, Houston Advanced Research Center (HARC),
 \\ Mitchell Campus, Woodlands, TX 77381, USA;\\ 
Academy of Athens, Division of Natural Sciences,
Athens 10679, Greece}\\[0.2cm]
{\em $^e$William I. Fine Theoretical Physics Institute, School of
 Physics and Astronomy, University of Minnesota, Minneapolis, MN 55455,
 USA}
 
 \end{center}

\vspace{0.1cm}
\centerline{\bf ABSTRACT}
\vspace{0.1cm}

{\small 
We develop a string-inspired model for particle cosmology, based on a
flipped SU(5)$\times$U(1) gauge group formulated in a no-scale supergravity
framework. The model realizes Starobinsky-like inflation, which we assume to be followed by strong reheating, with the GUT symmetry being
broken subsequently by a light `flaton' field whose decay generates a second stage of reheating.
We discuss the production of gravitinos and the non-thermal contribution made by their decays to the density of
cold dark matter, which is assumed to be provided by the lightest neutralino. We also discuss the
masses of light and heavy neutrinos and leptogenesis. As discussed previously~\cite{egnno4}, a key r\^ole is played by a superpotential
coupling between the inflaton, matter and GUT Higgs fields, called $\lambda_6$. We scan over possible values of $\lambda_6$,
exploring the correlations between the possible values of observables. We emphasize that the release
of entropy during the GUT transition allows large regions of supersymmetry-breaking parameter space that would otherwise
lead to severe overdensity of dark matter. Furthermore, we find that the Big Bang nucleosynthesis lower limit on
the reheating temperature of $\sim 1$~MeV restricts the supersymmetry-breaking scale to a range
${\cal O}(10)$~TeV that is consistent with the absence of supersymmetric particles at the LHC.
}

\vspace{0.05in}

\begin{flushleft}
October 2019
\end{flushleft}
\medskip
\noindent

\newpage

%%%%%%%%%%%%%%%%%%%%%%%%%%%
\section{Introduction}
%%%%%%%%%%%%%%%%%%%%%%%%%%%

There are many aspects of cosmology where particle physics is called upon to play key r\^oles. In decreasing order (roughly) of energy and temperature scale, these include cosmological inflation and the subsequent reheating, baryogenesis, the decoupling of cold dark matter (CDM) from matter made of Standard Model (SM) particles, Big-Bang
Nucleosynthesis (BBN), relic left-handed neutrinos and the cosmological constant (dark energy). In a previous paper~\cite{egnno4} we proposed a model approach to these issues in particle cosmology that we develop further in this paper.

The approach we follow is guided by the expectation that string theory is the underlying fundamental quantum `theory of everything' including gravity as well as the SM. For that reason, we adopt a theoretical framework for sub-Planckian physics that has been shown to be obtainable in principle from string theory. We expect that string theory is compactified on a manifold that preserves supersymmetry (SUSY) in the effective low-energy theory. The appropriate framework for combining SUSY with gravity is supergravity, of which string compactification picks out~\cite{Witten} the specific no-scale variety~\cite{no-scale,ekn2,LN}. 

As for the sub-Planckian gauge group, in weakly-coupled heterotic string compactifications the matter representations are limited in size, e.g., to $\mathbf{\overline{5}}$ and $\mathbf{10}$ representations of SU(5)~\footnote{There are more possibilities in strongly-coupled string constructions.}.  This consideration motivates our choice of GUT gauge group, namely flipped $\text{SU}(5) \times \text{U}(1)$~\cite{Barr,DKN,flipped2,AEHN}, which can be broken down to the SM $\text{SU}(3)_c \times \text{SU}(2)_L \times \text{U}(1)_Y$ group by a combination of $\mathbf{\overline{10}}$ and $\mathbf{10}$ Higgs representations, whereas conventional SU(5) and larger GUT groups require adjoint or larger Higgs representations.

We have explored different aspects of the combined no-scale flipped $\text{SU}(5) \times \text{U}(1)$ framework in a series of papers~\cite{egnno2,egnno3}, culminating in~\cite{egnno4}, where we analyzed possible resolutions of  many of the above-mentioned issues in particle cosmology. We emphasized there the key r\^ole played by one specific Yukawa coupling, denoted by $\lambda_6$, which connects $\mathbf{\overline{10}}$ matter, $\mathbf{{10}}$ Higgs and singlet inflaton fields. In this paper we provide more details of this no-scale flipped $\text{SU}(5) \times \text{U}(1)$ model, showing how the scale of SUSY breaking must be constrained for the consistency of this scenario for particle cosmology.

\begin{figure}[ht]
\centering
    \includegraphics[width=0.85\textwidth]{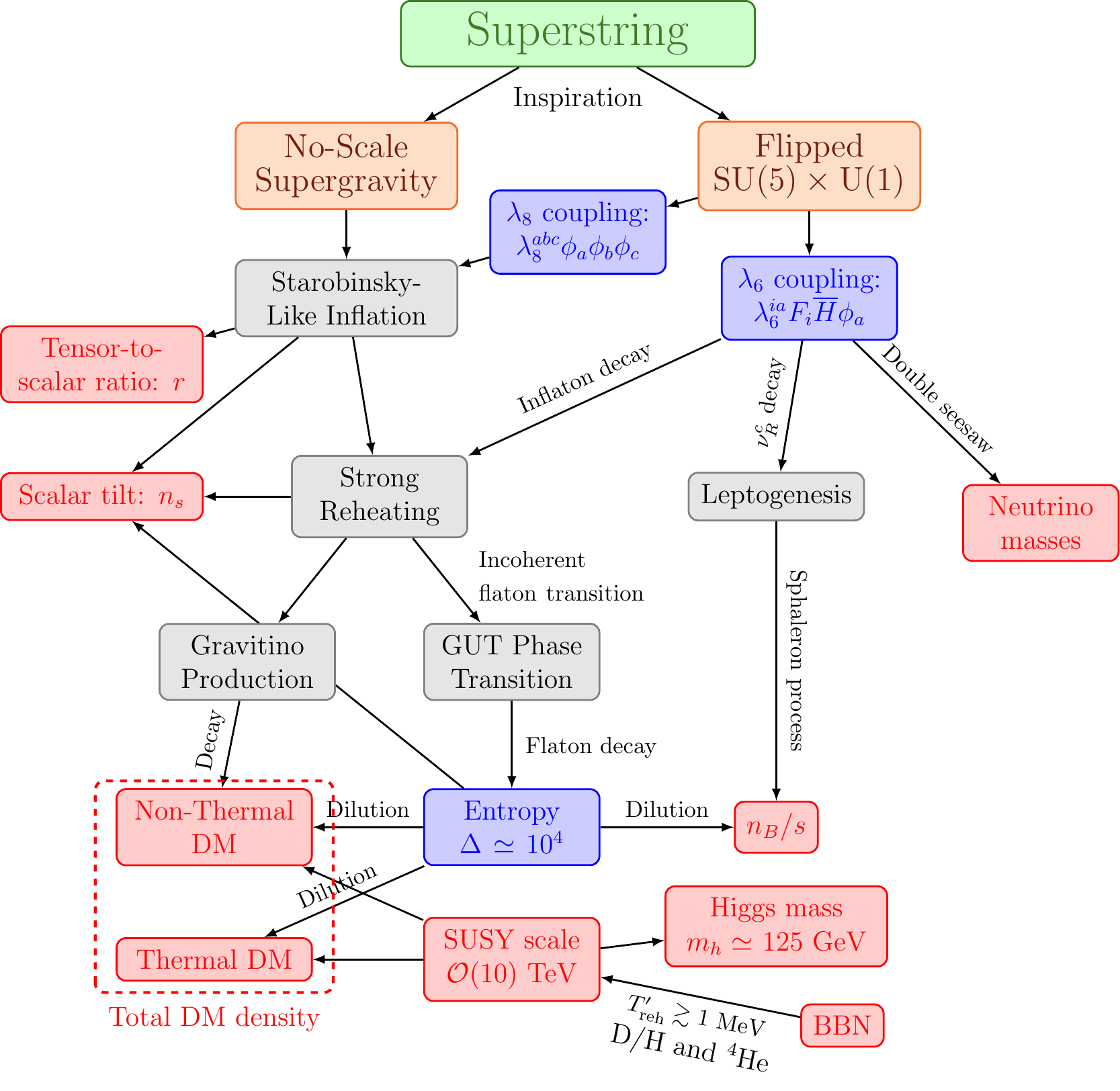}
%    \vspace{-3cm}
    \caption{\it The general structure of our scenario for particle cosmology.}
    \label{fig:concept}
\end{figure}

The general structure of our scenario for particle cosmology is illustrated in Fig.~\ref{fig:concept}. Inspired by superstring compactification models (as highlighted in green), we postulate no-scale supergravity and a flipped $\text{SU}(5) \times \text{U}(1)$ GUT. The latter includes a coupling $\lambda_8$ (highlighted in blue) that generates Starobinsky-like inflation with a successful prediction for the scalar spectral tilt, $n_s$, as well as a testable prediction for the tensor-to-scalar perturbation ratio, $r$. There is also a coupling $\lambda_6$ (also highlighted in blue) that plays key r\^oles in post-inflationary reheating, neutrino masses and leptogenesis as highlighted in~\cite{egnno4}. We postulate strong reheating, which leads to copious production of gravitinos that decay subsequently into non-thermal dark matter. The GUT $\text{SU}(5) \times \text{U}(1)$ $\to$ SM phase transition occurs after reheating, and generates a substantial amount of entropy, $\Delta$. Entropy dilution by a factor $\Delta \sim 10^4$ (also highlighted in blue) reduces the cosmological baryon asymmetry to the measured value, and also reduces the density of non-thermal and thermal dark matter so as to be compatible with Planck \cite{Aghanim:2018eyx}. These requirements and the lower limit of ${\cal O}({\rm MeV})$ on the reheating temperature after the GUT transition imposed by the success of conventional BBN \cite{bbn} prefer a SUSY breaking scale that is ${\cal O}(10)$~TeV. This and other key model predictions ($r$, $n_s$, neutrino masses, $n_B/s$, the dark matter density, the SUSY scale, BBN and the Higgs mass $m_h$) are highlighted in red in Fig.~\ref{fig:concept}.

%%%%%%%%%%%%%%%%%%%%%%%%%%%%%%%%%%%%%%%%%%%%%
%%%%%%%%%%%%%%%%%%%%%%%%%%%%%%%%%%%%%%%%%%%%%%%

The layout of this paper is as follows. In Section~\ref{sec:model} we review the construction of our model, reviewing the assignments of matter particles to $\text{SU}(5) \times \text{U}(1)$ representations and the singlet inflaton and flaton fields, and highlighting the importance of the $\lambda_6$ coupling. Then, in Section~\ref{sec:phasetrans} we review some cosmological aspects of our model, focusing on the reheating epoch following inflation, which we assume to be strong, and the subsequent breaking of the GUT symmetry via thermal corrections to the effective potential for the flaton. The amount of entropy, $\Delta$, generated during the transition to the SM gauge group is an important aspect of our analysis. As we discuss in Section~\ref{sec:gravitino}, strong reheating implies the copious production of gravitinos, which decay subsequently to CDM particles, assumed to be neutralinos. Their density may be reduced into the range that is cosmologically acceptable by the entropy factor $\Delta$.  We consider neutrino physics in Section~\ref{eq:neutrinomass}, performing a scan over values of the $\lambda_6$ coupling and studying possible values of the light neutrino masses, the baryon asymmetry generated via leptogenesis and the non-thermal contribution to the CDM density in Section~\ref{sec:scan}. As we discuss in Section~\ref{sec:pheno}, the dominant contribution to the CDM density may be thermal, which could be brought into the cosmological range for $\Delta \gg 1$ even for sparticle masses $\gg 1$~TeV. An important constraint is that the universe should reheat to a temperature $\gtrsim 1$~MeV, so that BBN can proceed successfully as in conventional Big Bang cosmology. This and the correct CDM density can be reconciled for $\Delta \sim 10^4$ and sparticle masses $\gtrsim 10$~TeV. This would explain why sparticles have not been detected at the LHC, whilst leaving open the possibility of detecting them at a future 100-TeV proton-proton collider such as FCC-hh~\cite{FCC-hh}. Finally, Section~\ref{sec:conclusion} summarizes our conclusions.

%%%%%%%%%%%%%%%%%%%%%%%%%%%%%%%%%%%%%%%%%%%%%%%%%%%%%%%%%%%%%%
\section{Model}
\label{sec:model}
%%%%%%%%%%%%%%%%%%%%%%%%%%%%%%%%%%%%%%%%%%%%%%%%%%%%%%%%%%%%%%

In the no-scale flipped ${\rm SU}(5) \times {\rm U}(1)$ GUT model \cite{Barr,DKN,flipped2,AEHN, egnno2, egnno3, egnno4}, the three generations of minimal supersymmetric Standard Model (MSSM) matter fields, together with three right-handed neutrino chiral superfields, are embedded into $\mathbf{10}$, $\bar{\mathbf{5}}$ and $\mathbf{1}$ representations. We denote the representations by $F_i$, $\bar{f}_i$, and $\ell^c_i$, respectively, with $i=1,2,3$ the generation index. A characteristic feature of the flipped ${\rm SU}(5) \times {\rm U}(1)$ GUT is that the assignments of the quantum numbers for right-handed leptons and the right-handed up- and down-type quarks are ``flipped'' with respect to the standard SU(5) assignments. In addition to these matter fields, this model contains a pair of $\mathbf{10}$ and $\overline{\mathbf{10}}$ Higgs fields, $H$ and $\bar{H}$, respectively; a pair of $\mathbf{5}$ and $\overline{\mathbf{5}}$ Higgs fields, $h$ and $\bar{h}$, respectively; and four singlet fields, $\phi_a$ ($a = 0, \dots, 3$). The $H$ and $\bar{H}$ fields break the ${\rm SU}(5) \times {\rm U}(1)$ gauge group down to the SM gauge group once these fields develop vacuum expectation values (VEVs). The phase transition associated with this symmetry breaking
was discussed in detail in \cite{egnno2,egnno3} and will be reviewed in Section~\ref{sec:phasetrans}. As in the MSSM, the $\mathrm{SU}(2)_L \times \mathrm{U}(1)_Y$ gauge symmetry is broken by the VEVs of the doublet Higgs fields $H_d$ and $H_u$, which reside in $h$ and $\bar{h}$, respectively. In Table~\ref{tab:model}, we summarize the field content and the charge assignments of the fields, where the U(1) charges are given in units of $1/\sqrt{40}$. For the notation of the component fields, we follow Ref.~\cite{flipped2}. With these charge assignments, the U(1)$_Y$ hypercharge $Y$ is given by the following linear combination of the SU(5) generator $T_{24} = \mathrm{diag}(2,2,2,-3,-3)/\sqrt{60}$ and the U(1) charge $Q_X$: 
\begin{equation}
    Y = \frac{1}{\sqrt{15}} T_{24} + \frac{1}{5} Q_X ~.
\end{equation}
In addition to the gauge symmetry, we assume that this model possesses an approximate $\mathbb{Z}_2$ symmetry which is respected at the renormalizable level but is violated by some Planck-scale suppressed operators. Only the $H$ field is odd under this $\mathbb{Z}_2$ symmetry, while the rest of the fields are even, as shown in the last column of Table~\ref{tab:model}. We expect that $\mathbb{Z}_2$-breaking non-renormalizable operators prevent the formation of domain walls when the field $H$ acquires a VEV. 

%%%%%%%%%%%%%%%% Table %%%%%%%%%%%%%%%%%%%%%%%%%%%%%%%%%%%%%%%%%%%%%%%%%%%%%%%%%%
\begin{table}[t]
\centering
%\vspace{3mm}
\begin{tabular}{l l ccc}
\hline \hline
Fields & Components & SU(5)\quad & U(1)\quad & $\mathbb{Z}_2$\quad\\
\hline
$F_i$ & $d^c_i$, $Q_i$, $\nu^c_i$ & $\mathbf{10}$ & $+1$ & $+$ \\ 
$\bar{f}_i$ & $u^c_i$, $L_i$ & $\overline{\mathbf{5}}$ & $-3$ & $+$\\ 
${\ell}_i^c$ & $e^c_i$ & ${\mathbf{1}}$ & $+5$& $+$ \\ 
$H$ & $d^c_H$, $Q_H$, $\nu^c_H$ & $\mathbf{10}$ & $+1$ & $-$ \\ 
$\bar{H}$ & ${d}^c_{\bar{H}}$, ${Q}_{\bar{H}}$, ${\nu}^c_{\bar{H}}$ & $\overline{\mathbf{10}}$ & $-1$ & $+$ \\ 
$h$ & $D$, $H_d$ & $\mathbf{5}$ & $-2$ & $+$ \\ 
$\bar{h}$ & $\bar{D}$, $H_u$ & $\overline{\mathbf{5}}$ & $+2$ & $+$\\ 
$\phi_a$ & $\phi_a$ & ${\mathbf{1}}$ & $0$& $+$ \\ 
 \hline
\hline
\end{tabular}
\caption{
\em The field content and the charge assignments in the flipped ${\rm SU}(5) \times {\rm U}(1)$ GUT model. The U(1) charges are given in units of $1/\sqrt{40}$. For the notation of the component fields, we follow Ref.~\cite{flipped2}.
}
\label{tab:model}
\end{table}
%%%%%%%%%%%%%%%%%%%%%%%%%%%%%%%%%%%%%%%%%%%%%%%%%%%%%%%%%%%%%%%%%%%%%%%%%%%%%%%%

The renormalizable superpotential in this model is then given by
\begin{align} \notag
W &=  \lambda_1^{ij} F_iF_jh + \lambda_2^{ij} F_i\bar{f}_j\bar{h} +
 \lambda_3^{ij}\bar{f}_i\ell^c_j h + \lambda_4 HHh + \lambda_5
 \bar{H}\bar{H}\bar{h}\\ 
&+ \lambda_6^{ia} F_i\bar{H}\phi_a + \lambda_7^a h\bar{h}\phi_a
 + \lambda_8^{abc}\phi_a\phi_b\phi_c + \mu^{ab}\phi_a\phi_b\,.
\label{Wgen} 
\end{align}
We note that the $\mathbb{Z}_2$ symmetry forbids some terms, such as $F_i H h$ and $\bar{f}_i H \bar{h}$, which are unwanted. After the field $H$ acquires a VEV, these terms would yield the operators $\langle H\rangle d_i^c D$ and $\langle H\rangle L_i H_u$, respectively, which induce baryon/lepton-number violation as well as $R$-parity violation. The $\mathbb{Z}_2$ symmetry also forbids a
vector-like mass term for $H$ and $\bar{H}$, which is advantageous for suppressing rapid proton decay induced by color-triplet Higgs exchange, as we discuss below. 

Since we work within a supergravity framework, we must specify the corresponding K\"ahler potential, which we assume to be of no-scale form \cite{ekn2}:
\begin{equation}
    K = -3 \ln \biggl[
    T + \bar{T} - \frac{1}{3} \sum_{\Psi} \left|\Psi \right|^2
    \biggr] ~,
\end{equation}
where $T$ is the volume modulus, and the sum over $\Psi$ includes all the chiral matter superfields in this model. In the absence of any moduli dependence of the gauge kinetic function, the scalar potential is 
\begin{equation}
    V = e^{2K/3} \biggl[
    \sum_\Psi \biggl|\frac{\partial W}{\partial \Psi}\biggr|^2 + \frac{1}{2} D^a D^a
    \biggr] ~,
\end{equation}
where the $D$-term part of the potential with vanishing SM
non-singlet fields is given by
\begin{equation}
 D^aD^a = \left(\frac{3}{10}g_5^2 +
\frac{1}{80} g_X^2\right)\,\left(|\tilde{\nu}^c_i|^2
+|\tilde{\nu}^c_{H}|^2-|\tilde{\nu}^c_{\bar{H}}|^2\right)^2~,
\end{equation}
where $g_5$ and $g_X$ are the coupling constants of the SU(5) and U(1)$_X$ gauge interactions, respectively. 
There is an $F$- and $D$-flat direction in the potential $V$, in the direction of a linear combination of $\tilde{\nu}_H^c$ and $\tilde{\nu}_{\bar H}^c$. We denote this combination by $\Phi$, and refer to it as the `flaton'. This flaton field is, therefore, massless in the supersymmetric limit~\footnote{We do not discuss in this paper the mechanisms for generating soft supersymmetry breaking or a cosmological constant (dark energy), which were considered recently in the no-scale context in~\cite{ENOV3}.}.

At low energies, the soft supersymmetry-breaking scalar mass-squared term for the flaton, $m^2_\Phi$, is driven negative by renormalization-group equation (RGE) effects due to the Yukawa couplings $\lambda_{4,5,6}$ \cite{flipped2}. This negative mass-squared term destabilizes the origin of the flat direction, and thus the flaton field develops a VEV, breaking the ${\rm SU}(5) \times {\rm U}(1)$ GUT symmetry into the SM gauge group. For a large field value of the flaton, the flat direction is uplifted by a Planck-scale-suppressed superpotential term of the form~\footnote{In general, we can consider operators of the form $(H\bar{H})^n/M_P^{2n-3}$. For $n < 4$, we need a very small coupling $\lambda$ and/or a very large flaton mass, in order to obtain a GUT-scale flaton VEV.}:
\begin{equation}
    W_{\text{NR}} \simeq \frac{\lambda}{M^{5}_{P}} (H \bar{H})^4 ~,
\end{equation}
where $M_P \equiv (8\pi G_N)^{-1/2}$ denotes the reduced Planck mass. In this case, there is a relation between the flaton VEV, $\langle \Phi \rangle$, and the soft supersymmetry-breaking mass term of the flaton:
\begin{equation}
    |m_\Phi| \simeq \frac{|\lambda| \langle \Phi \rangle^6}{M_P^5}
    \simeq |\lambda| \times \biggl(\frac{\langle \Phi \rangle}{10^{16}~{\rm GeV}}\biggr)^6 \times 12 ~{\rm TeV}~.
\end{equation}
 After $\Phi$ acquires a VEV, 13 gauge vector multiplets among the 25 in the ${\rm SU}(5) \times {\rm U}(1)$ gauge group become massive by absorbing the corresponding Nambu-Goldstone (NG) chiral superfields in $H$ and $\bar{H}$. Besides these 13 NG fields and the flaton $\Phi$, there are six components in $H$ and $\bar{H}$, $d_H^c$ and $d_{\bar{H}}^c$. These fields form vector-like multiplets with $D$ and $\bar{D}$, acquiring masses of order $\lambda_4 \langle \Phi \rangle$ and $\lambda_5 \langle \Phi \rangle$, respectively. On the other hand, the electroweak doublets $H_d$ and $H_u$ in $h$ and $\bar{h}$ do {\it not} acquire masses from the flaton VEV---this is an economical realization of the missing-partner mechanism \cite{flipped2, Masiero:1982fe} that solves naturally the doublet-triplet splitting problem. As a result, the color-triplet Higgs fields $D$ and $\bar{D}$ become massive despite the absence of a vector-like mass term $h\bar{h}$.  

The exchange of the color-triplet Higgs fields in general induces dimension-five baryon- and lepton-number violating operators \cite{Sakai:1981pk}, which cause rapid proton decay. In the present setup, however, such operators are extremely suppressed; in order for these operators to be induced via color-triplet Higgs exchange, a chirality flip due to vector-like mass terms, $D\bar{D}$ or $d_H^c d_{\bar{H}}^c$, would be required, but these terms are absent in Eq.~\eqref{Wgen}.\par\bigskip

As discussed in detail in Ref.~\cite{egnno2}, this model offers a successful framework for inflation, where one of the singlet fields plays the role of the inflaton. We call it $\phi_0$ in the following discussion. It is then found that if 
\begin{equation}
    \mu^{00} = \frac{1}{2}m_s~, \qquad \lambda_8^{000} = -\frac{m_s}{3 \sqrt{3} M_P}~,
    \label{eq:starobinskycond}
\end{equation}
an asymptotically-flat Starobinsky-like potential \cite{Staro} is obtained for $\phi_0$ \cite{ENO6}, and for $m_s \simeq 3 \times 10^{13}$~GeV, the measured amplitude of the primordial power spectrum  is successfully reproduced. Since the potential is Starobinsky-like, the tensor-to-scalar ratio $r \simeq 3 \times 10^{-3}$, well within the range allowed by the Planck and other data~\cite{Aghanim:2018eyx}. This prediction can be tested in future CMB experiments such as CMB-S4 \cite{Abazajian:2016yjj} and LiteBIRD \cite{Hazumi:2019lys}. Additionally, as we discuss later, the predicted value of the tilt in the scalar perturbation spectrum, $n_s$ is also within the range favoured by Planck and other data at the 68\% CL. This is the inflationary scenario we consider in this paper~\footnote{We note also that, as discussed in~\cite{ENOV3}, a dark energy term may easily be added to the model.}.

As seen in Eq.~\eqref{Wgen}, the inflaton $\phi_0$ can couple to the matter sector via the couplings $\lambda_6$ and $\lambda_7$. In Ref.~\cite{egnno2}, two distinct cases, $\lambda_6^{i0} = 0$ (Scenario A) or $\lambda_6^{i0} \neq 0$ (Scenario B), were separately studied. We focus on Scenario B in this work. In this scenario, one of the three singlet fields other than $\phi_0$, which we denote by $\phi_3$, does not have the $\lambda_6$ coupling; i.e., $\lambda_6^{i3} = 0$, whereas $\lambda_6^{ia} \neq 0$ for $i = 1,2,3$ and $a = 0,1,2$. We also assume $\lambda_7^a = 0$ for $a = 0,1,2$. To realize this scenario, we introduce a modified $R$-parity, under which the fields in this model transform as
\begin{align}
    F_i, \bar{f}_i, \ell_i^c, \phi_0, \phi_1, \phi_2
    &\to 
    -F_i, -\bar{f}_i, -\ell_i^c, -\phi_0, -\phi_1, -\phi_2 ~, \nonumber \\[3pt]
    H, \bar{H}, h, \bar{h}, \phi_3 &\to 
    H, \bar{H}, h, \bar{h}, \phi_3 ~.
\end{align}
We note that this modified $R$-parity is slightly violated by the coupling $\lambda_8^{000}$ in Eq.~\eqref{eq:starobinskycond}~\footnote{The violation of the modified $R$-parity may be evaded if we use a higher-dimensional quartic superpotential term for the inflaton potential, instead of the trilinear coupling $\lambda_8^{000}$. Indeed, as discussed in Ref.~\cite{Khalil:2018iip}, a potential with
Starobinsky-like properties can be obtained from a superpotential that consists only of quadratic and quartic terms for the inflaton. In this case, the inflaton can be $R$-parity odd without violating $R$-parity.}. Nevertheless, since this $R$-parity-violating effect is only very weakly transmitted to the matter sector, the lifetime of the lightest supersymmetric particle (LSP) is still much longer than the age of the Universe \cite{egnno3, ENO8}, so the LSP can be a good dark matter candidate. We also note that the singlet $\phi_3$ can acquire a VEV without spontaneously breaking the modified $R$-parity. In this case, the coupling $\lambda_7^3$, which is allowed by the modified $R$-parity, generates an effective $\mu$ term for $H_u$ and $H_d$, $\mu = \lambda_7^3 \langle \phi_3 \rangle$, just as in the next-to-minimal supersymmetric extension of the SM.

%%%%%%%%%%%%%%%%%%%%%%%%%%%%%%%%%%%%%%%%%%%%%
\section{Reheating and the GUT Phase Transition}
\label{sec:phasetrans}
%%%%%%%%%%%%%%%%%%%%%%%%%%%%%%%%%%%%%%%%%%%%

We discuss now the aftermath of inflation, focusing on the portion of the model
parameter space where the {strong reheating} scenario discussed
in Ref.~\cite{egnno3} is realized. As shown in Ref.~\cite{egnno3}, in this case the
GUT symmetry is left unbroken at the end of inflation, and we further assume
that the system remains in the unbroken phase during reheating, 
as is confirmed in the following analysis. 
The GUT phase transition is triggered by the difference in the number of
light degrees of freedom, $g$, between the broken and unbroken phases~\cite{supercosm,NOT,Campbell:1987eb, egnno2, egnno3}. Massless
superfields provide a thermal correction to the effective potential of $-
g\pi^2 T^4/90$, where $T$ denotes the temperature of the Universe. Since
the number of light degrees of freedom in the unbroken phase ($g = 103$)
is larger than that in the Higgs phase ($g = 62$), $\Phi$ is kept at the
origin at high temperatures. However, once the temperature drops below
 the confinement scale of the SU(5) gauge theory, $\Lambda_c$,
the number of light degrees of freedom significantly decreases ($g \leq
25$), and thus the Higgs phase becomes energetically favored
\cite{egnno2}. We have found that in this strong reheating scenario the incoherent component of the
flaton drives the phase transition if $\Lambda_c \gtrsim 2.3 (m_\Phi
M_{\rm GUT})^{1/2}$ \cite{egnno3}, where $m_\Phi$ and $M_{\rm GUT}$ are
the flaton mass and the GUT scale, respectively. For $m_\Phi = 10^4$~GeV and
$M_{\rm GUT} =10^{16}$~GeV, the above condition leads to the requirement $\Lambda_c
\gtrsim 2.3 \times 10^{10}$~GeV. Note that even if the GUT phase transition occurs after inflation ends, our model does not suffer from the monopole problem, contrary to the conventional SU(5) GUTs, since our model is based on the product group $\text{SU}(5) \times \text{U}(1)$.

It was shown in Ref.~\cite{egnno3} that if $|\lambda_{6}^{i0}| \gtrsim
{\cal O}(10^{-4})$ reheating is completed in the symmetric phase
via the dominant inflaton decay channel $\phi_0 \to F_i \bar{H}$. The reheating
temperature in this case is given by
\begin{equation}
 T_{\rm reh} \simeq 1.7 \times 10^{15} ~\mathrm{GeV}
\times \sqrt{\sum_{i} |\lambda_6^{i0}|^2} ~,
\label{eq:treh}
\end{equation}
indicating a direct relation between
$T_{\rm reh}$ and $\lambda_6$.

In the case of such strong reheating, 
the flaton decouples
from the thermal bath, and when $T \lesssim m_\Phi$ it becomes
non-relativistic and eventually dominates the energy density of the Universe until
it decays, generating a second episode of
reheating. 
The amount of entropy
released by flaton decay is estimated to be
\begin{equation}\label{eq:delta}
 \Delta \simeq 1.6 \times 10^4\, \lambda^{-2}_{1,2,3,7} \,
\biggl(\frac{M_{\rm GUT}}{10^{16}~{\rm GeV}}\biggr)
\biggl(\frac{10~{\rm TeV}}{m^2_{\rm soft}/m_\Phi}\biggr)^{1/2} ~,
\end{equation} 
where $m_{\rm soft}$ stands for a typical sfermion
mass. As we see later, values of $\Delta \gg 1$ are favoured in this model.

%%%%%%%%%%%%%%%%%%%%%%%%%%%%%%%%%%%%%%
\section{Gravitino Production}
\label{sec:gravitino}
%%%%%%%%%%%%%%%%%%%%%%%%%%%%%%%%%%%

Gravitinos are produced during reheating via the
scattering/decay of particles in the thermal bath~\cite{weinberg,elinn,nos,ehnos,kl,ekn,eln,Juszkiewicz:gg,mmy,Kawasaki:1994af,
Moroi:1995fs,enor,Giudice:1999am,bbb,kmy,stef,Pradler:2006qh,ps2,rs,kkmy,
EGNOP}. For
the calculation of the gravitino production rate, we use the formalism 
outlined in \cite{rs}, but with the 
group-theoretical factors and couplings
appropriate to flipped $\text{SU}(5) \times \text{U}(1)$. In this case, the total gauge contribution to the thermally-averaged cross section for gravitino production can be written as
\beq\label{eq:sigv}
\langle \sigma v\rangle_{\rm gauge} \;\simeq\; \frac{1}{8\pi \zeta(3)^2 M_P^2} \sum_{i} \left[ 1.29 c_i g_i^2 + \frac{\pi^2}{2} n_i f_i\left(\frac{m_{V_i}}{T}\right) \right] \left(1 + \frac{M_i^2}{3 m_{3/2}^2}\right)\,,
\eeq 
where $m_{3/2}$ denotes the gravitino mass, $g_i$ and $M_i$ are the gauge coupling constant and gaugino mass, respectively, and 
the sum is taken over the corresponding gauge groups: $\text{SU}(5)\times \text{U}(1)_X$  in the unbroken phase and $\text{SU}(3)_c \times \text{SU}(2)_L \times \text{U}(1)_Y$ in the broken phase. The group-theoretical factors $c_i$ and $n_i$ are related to the Casimirs of the corresponding matter and gauge representations, and the number of corresponding vector fields, respectively, and can be found in Table~\ref{tab:res}. The values of the thermal vector masses $m_{V_i}$ can also be found in Table~\ref{tab:res}. The quasi-universal rate functions $f_i$ encode the one-loop thermal effects relevant to the process, and their derivation and computation beyond the hard-thermal-loop approximation in the broken phase can be found in~\cite{rs}. In the unbroken phase we content ourselves with extrapolating the rate function for SU(3)$_c$ to the case of the $\text{SU}(5) \times \text{U}(1)_X$ group, which leads to a $\lesssim \mathcal{O}(1)$ error in our approximation. Finally, we note that the first term in the gaugino mass-dependent factors $(1 + M_i^2/3 m_{3/2}^2 )$ corresponds to the production of the transversely polarized gravitino, while the second term is associated with the production of the longitudinal (Goldstino) component.

\begin{table}[!ht]
\centering
{ % begin box to localize effect of arraystretch change
\renewcommand{\arraystretch}{1.4}
\begin{tabular}{c c c c}
\hline \hline
Gauge group & $c_i$ & $n_i$ & $m_{V_i}^2/T^2$ \\
\hline
SU(3)$_c$ & 24 & 8 & $\frac{9}{4}g_3^2$ \\ 
SU(2)$_L$ & 15 & 3 & $\frac{9}{4}g_2^2$ \\ 
U(1)$_Y$ & 11 & 1 & $\frac{11}{4}g_1^2$ \\ 
\hline
SU(5) & 120 & 24 & $\frac{15}{4}g_5^2$ \\
U(1)$_X$ & $\frac{15}{2}$ & 1 & $\frac{15}{8}g_X^2$ \\
 \hline
\hline
\end{tabular}}
\caption{\em Values of the gauge-group-dependent constants $c_i$, $n_i$, and the thermal vector masses $m_{V_i}$ in the parametrization (\ref{eq:sigv}) for the total gravitino cross section. Top: MSSM. Bottom: flipped ${\rm SU}(5) \times {\rm U}(1)$.}
\label{tab:res}
\end{table}

\begin{figure}[ht]
\centering
    \includegraphics[width=0.89\textwidth]{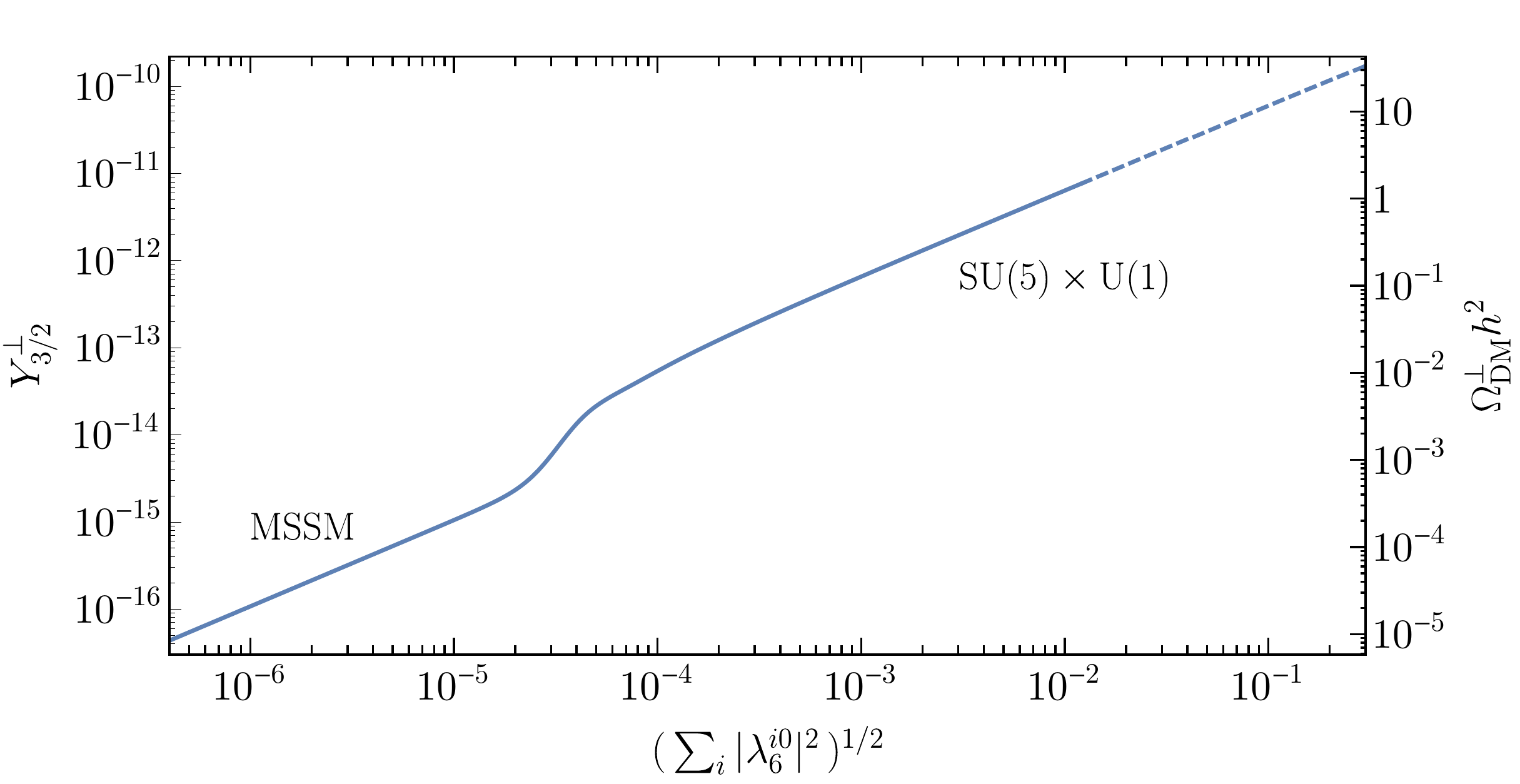}
       \caption{\it The transverse gravitino yield at $T \ll 1\,{\rm MeV}$ as a function of the effective Yukawa coupling
assuming strong reheating. The left vertical axis corresponds to the numerically-calculated yield including the dilution factor $\Delta$ given by (\ref{eq:delta}). The right vertical axis shows the corresponding DM closure fraction, assuming $m_{\rm LSP} = 10\,{\rm TeV}$. The dashed portion of the yield curve corresponds to Yukawa couplings for which in-medium and non-perturbative effects may have an impact on the inflaton decay rate. }
    \label{fig:gravitino}
\end{figure}

Numerical integration of the Boltzmann equation 
\beq
\frac{dn_{3/2}}{dt} + 3Hn_{3/2} \;=\; \langle \sigma v\rangle n_{\rm rad}^2
\eeq
for the transverse gravitino yield leads to the results displayed in Fig.~\ref{fig:gravitino}. Here $n_{\rm rad}=\zeta(3)T^3/\pi^2$ is the number density of any single bosonic relativistic degree of freedom. The numerical results are provided in terms of the gravitino yield
\beq
Y_{3/2} \;\equiv\; \frac{n_{3/2}}{n_{\rm rad}}\,, 
\eeq
evaluated at low temperatures $T\ll 1\,{\rm MeV}$, and account for the running of the gauge couplings at one loop. These results are an improvement over those presented in~\cite{egnno3}, which ignored the increase in the number of degrees of freedom in the plasma in the unbroken GUT phase. 

The gravitinos produced during reheating eventually decay into
LSPs. Fitting the numerical results for gravitino production, we find the following non-thermal, transverse contribution to the LSP abundance in the symmetric phase:
\begin{align}
 \Omega_{\rm DM} h^2\bigr|_{\text{non-thermal}} \;&\simeq\;  0.12  \,
\biggl(\frac{1.6 \times 10^4}{\Delta}\biggr)
\biggl(\frac{m_{\rm LSP}}{10~{\rm TeV}}\biggr)
\biggl(\frac{\sqrt{\sum_{i} |\lambda^{i0}_6|^2}}{0.97 \times 10^{-3}}\biggr) \nonumber \\
 &=\;  0.12  \,
\biggl(\frac{1.6 \times 10^4}{\Delta}\biggr)
\biggl(\frac{m_{\rm LSP}}{10~{\rm TeV}}\biggr)
\biggl(\frac{T_{\rm reh}}{1.6 \times 10^{12} ~{\rm GeV}}\biggr)
~.
\label{eq:odm}
\end{align}
Thus the LSP relic density
is also directly related to $\lambda_6$. The total dark matter abundance is obtained by adding this non-thermal
component to the thermal relic density of the LSP,
and is reduced by a dilution factor of
$\Delta$.

%%%%%%%%%%%%%%%%%%%%%%%%%%%%%%%%%%%%%%%%%%
\section{Neutrino Masses and Leptogenesis}
\label{eq:neutrinomass}
%%%%%%%%%%%%%%%%%%%%%%%%%%%%%%%%%%%%%%%%%%

The $\lambda_6$ coupling in this model also plays an important role in determining the neutrino mass structure \cite{egnno2, egnno3, egnno4}. In this Section, we study this structure in detail, following Ref.~\cite{Ellis:1993ks}. 
We adopt the basis where $\lambda_2^{ij}$ and $\mu^{ab}$ are real and diagonal without loss of generality \footnote{This basis corresponds to the case where $U_u = U_{u^c} = U_\phi = \1$ in Ref.~\cite{Ellis:1993ks}. }. In this case, the diagonal components of these matrices are given by
\begin{equation}
 \lambda_2 \simeq \frac{1}{\langle \bar{h}_0 \rangle}
{\rm diag}(m_u, m_c, m_t) ~, \qquad
 \mu = \frac{1}{2}{\rm diag} (m_s, \mu^1, \mu^2) ~,
 \label{eq:lam2andmu}
\end{equation}
where we take $m_s = 3\times 10^{13}$~GeV (see Section~\ref{sec:model}). We 
express these matrices as  $\lambda_2^{ij} = \lambda_2^i \delta^{ij}$ and $\mu^{ab} = \mu^a \delta^{ab}$ in what follows. The first equation in Eq.~\eqref{eq:lam2andmu} is only an approximate expression; in general, renormalization-group effects and threshold corrections cause $\lambda_2$ to deviate from the up-type Yukawa couplings at low energies by at most ${\cal O}(10)\%$. Since these effects depend on the mass spectrum of the theory, we neglect such corrections in the following analysis.

The superpotential terms relevant to the
present discussion are 
\begin{equation}
 W = \sum_{i=1}^{3} \lambda_{2}^i \nu_i^c L_i H_d 
+ \sum_{a=0}^{2} \mu^a \phi_a^2 
+ \sum_{i,a} \lambda_6^{ia} \nu_i^c \nu_{\bar{H}}^c \tilde{\phi}_a ~,
\end{equation}
where $\lambda_6^{ia}$ is a $3\times 3$ complex
matrix \footnote{The coupling $\lambda_6$ defined here is the same as $\lambda_6^\prime$ in
Ref.~\cite{Ellis:1993ks}, but this difference does not matter since we
will consider a generic complex matrix for $\lambda_6$.}.
As we noted above, only three singlet fields, including the inflaton, couple
to the neutrino sector.
The neutrino/singlet-fermion mass matrix can be written as 
\begin{equation}
 {\cal L}_{\rm mass}=
-\frac{1}{2}
\left(
\begin{matrix}
{\nu}_{i} & {\nu}_{j}^c & \tilde{\phi_a}
\end{matrix}
\right) 
\begin{pmatrix}
 0 & \lambda_2^{i j}\langle \bar{h}_0\rangle & 0 \\
\lambda_2^{i j}\langle \bar{h}_0\rangle & 0 &
 \lambda_6^{j a}\langle \tilde{\nu}_{\bar{H}}^c\rangle \\
0 & \lambda_6^{j a}\langle \tilde{\nu}_{\bar{H}}^c\rangle & \mu^a
\end{pmatrix}
 \left(
\begin{matrix}
\nu_{i} \\ \nu_{j}^c \\ \tilde{\phi_a}
\end{matrix}
\right)
+ {\rm h.c.}~,
\end{equation}
where $i,j = 1,2,3$ and $a= 0,1,2$, as before,
and $\tilde{\phi}_0$ corresponds to the fermionic superpartner of the inflaton field $\phi_0$. 
The mass matrix of the right-handed
neutrinos is obtained from a first seesaw mechanism:
\begin{equation}
 (m_{\nu^c})_{ij} = \sum_{a=0,1,2} \frac{\lambda_6^{ia} \lambda_6^{ja}}{\mu^a}
  \langle \tilde{\nu}_{\bar{H}}^c \rangle^2 ~,
\label{eq:mnuc}
\end{equation}
where we take $\langle \tilde{\nu}_{\bar{H}}^c \rangle = 10^{16}$~GeV in this
paper. We diagonalize the mass matrix (\ref{eq:mnuc})
using a unitary matrix $U_{\nu^c}$:
\begin{equation}
     m_{\nu^c}^D = U_{\nu^c}^T m_{\nu^c}
U_{\nu^c}  ~.
\label{eq:mnucdiagonalization}
\end{equation}
The light neutrino mass matrix is then obtained through a second
seesaw mechanism~\cite{Minkowski:1977sc,Georgi:1979dq}:
\begin{equation}
 (m_\nu)_{ij} = \sum_{k} \frac{\lambda_2^i \lambda_2^j (U_{\nu^c})_{ik}
  (U_{\nu^c})_{jk} \langle \bar{h}_0 \rangle^2 }{(m_{\nu^c}^D)_k} ~.
\label{eq:mnu}
\end{equation}
This mass matrix is diagonalized by a unitary matrix $U_\nu$, so that 
\begin{equation}
    m_\nu^D
= U_\nu^* m_\nu  U_\nu^\dagger ~.
\label{eq:mnudiagonalization}
\end{equation}
We note that, given a matrix
$\lambda^{ia}_6$, the eigenvalues of the $m_\nu$ and $m_{\nu^c}$ matrices, as well as the mixing matrices $U_{\nu^c}$ and $U_\nu$, are uniquely
determined as functions of $\mu^1$ and $\mu^2$ 
via Eqs.~(\ref{eq:mnuc}--\ref{eq:mnu}).

As can be seen from Eq.~\eqref{eq:mnuc}, the masses of the right-handed neutrinos are generated after $\bar{H}$ acquires a VEV, namely, after the GUT phase transition has completed. Therefore, in the strong reheating
scenario described in Section~\ref{sec:phasetrans}, the right-handed neutrinos are massless and in thermal equilibrium right after reheating is completed. When the GUT phase transition occurs, they obtain masses, and decouple from the thermal bath almost immediately if their masses are larger than the transition temperature, which we will confirm in the next Section. These right-handed neutrinos (and sneutrinos) subsequently decay into $L_i$ and $H_d$ non-thermally \cite{NOT, egnno3} \footnote{As we see in the next section, singlet fermions $\tilde{\phi}_a$ have masses well above the transition temperature, and thus the contribution of these fields to the generation of lepton asymmetry can safely be neglected. },  generating a lepton asymmetry \cite{fy} \footnote{We note that this lepton asymmetry is not washed out by the dimension-five operators of the form $L_i L_j \bar{h}\bar{h}$ that are obtained by integrating out the right-handed neutrinos and singlet fields. Since these operators are generated in the same manner as the neutrino mass matrix in Eq.~\eqref{eq:mnu}, in the bases where the neutrino mass matrix is diagonalized, the dimension-five operators can be written as $m_i L_iL_i \bar{h}\bar{h}/\langle \bar{h}_0\rangle^2 $, with $m_i$ ($i=1,2,3$) the mass eigenvalues of the light neutrinos. These operators decouple from the thermal bath at the time of the GUT phase transition for $m_i \lesssim 0.2$~eV, which is the case in our scenario, as we see in the next Section.  }. The sphaleron process \cite{Kuzmin:1985mm,cdo} then converts this lepton asymmetry partially to a baryon asymmetry. The resultant amount of the baryon asymmetry is given by
\begin{equation}
 \frac{n_B}{s} = -\frac{28}{79} \cdot \frac{135 \zeta (3)}{4\pi^4 g_{\rm
  reh}
  \Delta } \sum_{i=1,2,3} \epsilon_i ~,
  \label{nbs}
\end{equation}
where \cite{Ellis:1993ks, egnno3}
\begin{equation}
 \epsilon_i 
= 
\frac{1}{2\pi}\frac{ \sum_{j\neq i} {\rm Im} 
\left[\left(
U_{\nu^c}^\dagger (\lambda_2^D)^2 U_{\nu^c}
\right)_{ji}^2\right] }{\left[
U_{\nu^c}^\dagger (\lambda_2^D)^2 U_{\nu^c}
\right]_{ii}}
g\biggl(\frac{m^2_{\nu^c_{j}}}{m^2_{\nu^c_{i}}}\biggr) ~,
\end{equation}
with \cite{luty,Covi:1996wh}
\begin{equation}
 g(x) \equiv -\sqrt{x} \biggl[
\frac{2}{x-1} + \ln \biggl(\frac{1+x}{x}\biggr)
\biggr] ~.
\label{eq:gx}
\end{equation}
 {\it We note that the sign in (\ref{nbs}) is 
 fixed}: we must require $\sum_{i} \epsilon_i < 0$ in order to obtain $n_B/s > 0$.
Both the sign and magnitude of $\epsilon_i$ depend on the CP phases in the unitary matrix $U_{\nu^c}$, which are related to the CP phases in the matrix $\lambda_6$ through Eqs.~\eqref{eq:mnuc} and \eqref{eq:mnucdiagonalization}. Notice that the matrix $U_{\nu^c}$ is {\it not} directly related with the Pontecorvo-Maki-Nakagawa-Sakata (PMNS) mixing matrix \cite{Pontecorvo:1957cp}, $U_{\rm PMNS}$. As discussed in Ref.~\cite{Ellis:1993ks}, the PMNS
matrix is given by
\begin{equation}
    U_{\rm PMNS} = U_l^{} U_\nu^\dagger ~,
    \label{eq:pmns}
\end{equation}
where $U_l$ is a unitary matrix that is used to diagonalize the matrix $\lambda_3^{ij}$, which leads to the charged lepton Yukawa couplings. Although the matrix $U_\nu$ in Eq.~\eqref{eq:pmns} does have a connection with $U_{\nu^c}$ via Eqs.~\eqref{eq:mnu} and \eqref{eq:mnudiagonalization}, due to the presence of an extra unknown matrix, $U_l$, we cannot predict the PMNS matrix in our model. In particular, the determination of neutrino CP phases at low energies has no implication on the CP phases relevant to leptogenesis (i.e., the phases in the matrix $U_{\nu^c}$) \footnote{Once the PMNS matrix is determined, however, we can use Eq.~\eqref{eq:pmns} to relate the matrices $U_l$ and $U_\nu$.  As discussed in Ref.~\cite{Ellis:1993ks}, the matrix $U_l$ affects proton decay branching fractions and makes them different from those predicted in the standard SU(5) GUT. A more detailed discussion on proton decay in the flipped $\text{SU}(5) \times \text{U}(1)$ model will be given on another occasion \cite{egnno6}.}.

%%%%%%%%%%%%%%%%%%%%%%%%%%%%%%%%%%%%%%%%%%%%%%
\section{Scan of the Model Parameter Space}
\label{sec:scan}
%%%%%%%%%%%%%%%%%%%%%%%%%%%%%%%%%%%%%%%%%%%%%%

As we have seen in Section~\ref{sec:phasetrans}, the coupling $\lambda_6$ determines the reheating temperature, which then fixes the non-thermal
component of the dark matter abundance as shown in Eq.~\eqref{eq:odm}. This coupling also controls the neutrino mass structure and baryon asymmetry as discussed in~\cite{egnno4} and the previous Section. We now investigate numerically the effect of the $\lambda_6$ coupling on these quantities by performing a parameter scan of $\lambda_6$. We write it in the form
\begin{equation}
  \lambda_6 = r_6 M_6 ~,  
\end{equation}
where $r_6$ is a real constant, which plays a role of a scale factor, and $M_6$ is a generic complex $3\times 3$ matrix. We then scan
$r_6$ with a logarithmic distribution over the range $(10^{-4},
1)$ choosing a total of 2000 values. For each value of $r_6$, we generate 2000 random complex $3\times 3$
matrices $M_6$ with each component taking a value of ${\cal O}(1)$. This is the same data set as that used in Ref.~\cite{egnno4}.

As discussed in Section~\ref{eq:neutrinomass}, for each $3\times 3$ matrix $\lambda_6$, the eigenvalues of the $m_\nu$ and $m_{\nu^c}$ matrices and the mixing matrices $U_{\nu^c}$ and $U_\nu$ are obtained as functions of $\mu^1$ and $\mu^2$. We then determine these two $\mu$ parameters by requiring that the observed values of the squared mass differences, $\Delta m_{21}^2 \equiv m_2^2 - m_1^2$ and $\Delta m_{3\ell }^2 \equiv m_3^2 - m_\ell^2$, are reproduced within the experimental uncertainties, where $\ell = 1$ for the normal ordering (NO) case and $\ell = 2$ for the inverted ordering (IO) case.  For the experimental input, we use the results given in Ref.~\cite{nufit}, which we summarize in Table~\ref{tab:input}. 
We generate the same number of $\lambda_6$ matrices for each mass
ordering, and find solutions for 9839 and 730 matrix choices for
the NO and IO cases, respectively,
out of a total of $4 \times 10^6$ models sampled. This difference indicates that the NO case is favored in our model. 

%%%%%%%%%%%%%%%% Table %%%%%%%%%%%%%%%%%%%%%%%%%%%%%%%%%%%%%%%%%%%%%%%%%%%%%%%%%%
\begin{table}[ht!]
\centering
%\vspace{3mm}
\begin{tabular}{l |c c|cc}
\hline \hline
& \multicolumn{2}{c|}{Normal Ordering} & \multicolumn{2}{c}{Inverted Ordering}\\
 & Best fit & 3$\sigma$ range &  Best fit & 3$\sigma$ range \\
\hline
$\Delta m_{21}^2$ [$10^{-5}~\text{eV}^2$] & 7.39 & 6.79--8.01& 7.39& 6.79--8.01\\
$\Delta m_{3\ell }^2$ [$10^{-3}~\text{eV}^2$] & $2.525$& $ 2.431$--$2.622$& $-2.512$& $-(2.413$--$2.606)$\\
 \hline
\hline
\end{tabular}
\caption{\em
Input values for the squared mass differences of active neutrinos \cite{nufit}. 
}
\label{tab:input}
\end{table}
%%%%%%%%%%%%%%%%%%%%%%%%%%%%%%%%%%%%%%%%%%%%%%%%%%%%%%%%%%%%%%%%%%%%%%%%%%%%%%%%

\begin{figure}[ht!]
\centering
    \includegraphics[width=0.45\textwidth]{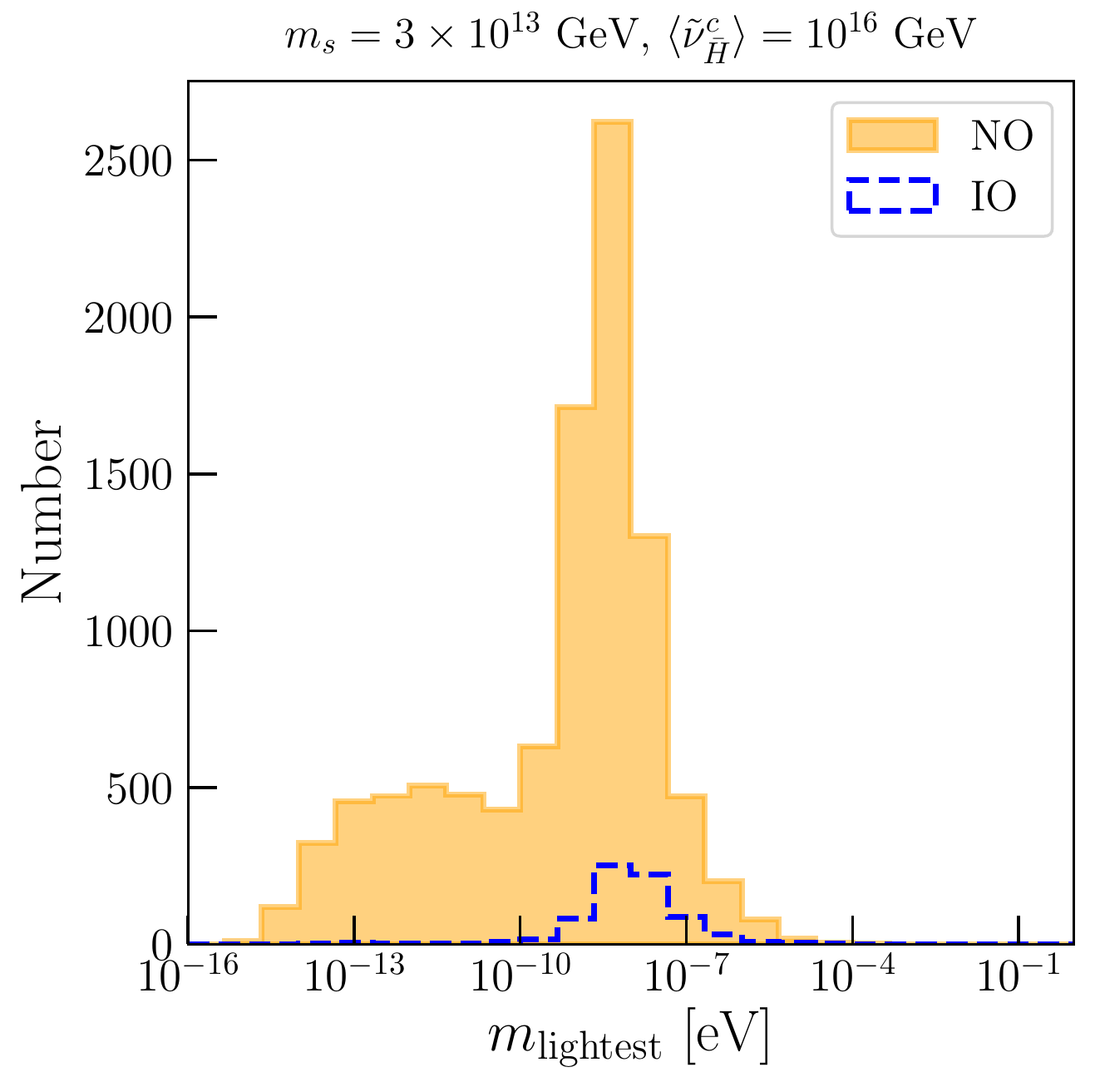}
       \includegraphics[width=0.47\textwidth]{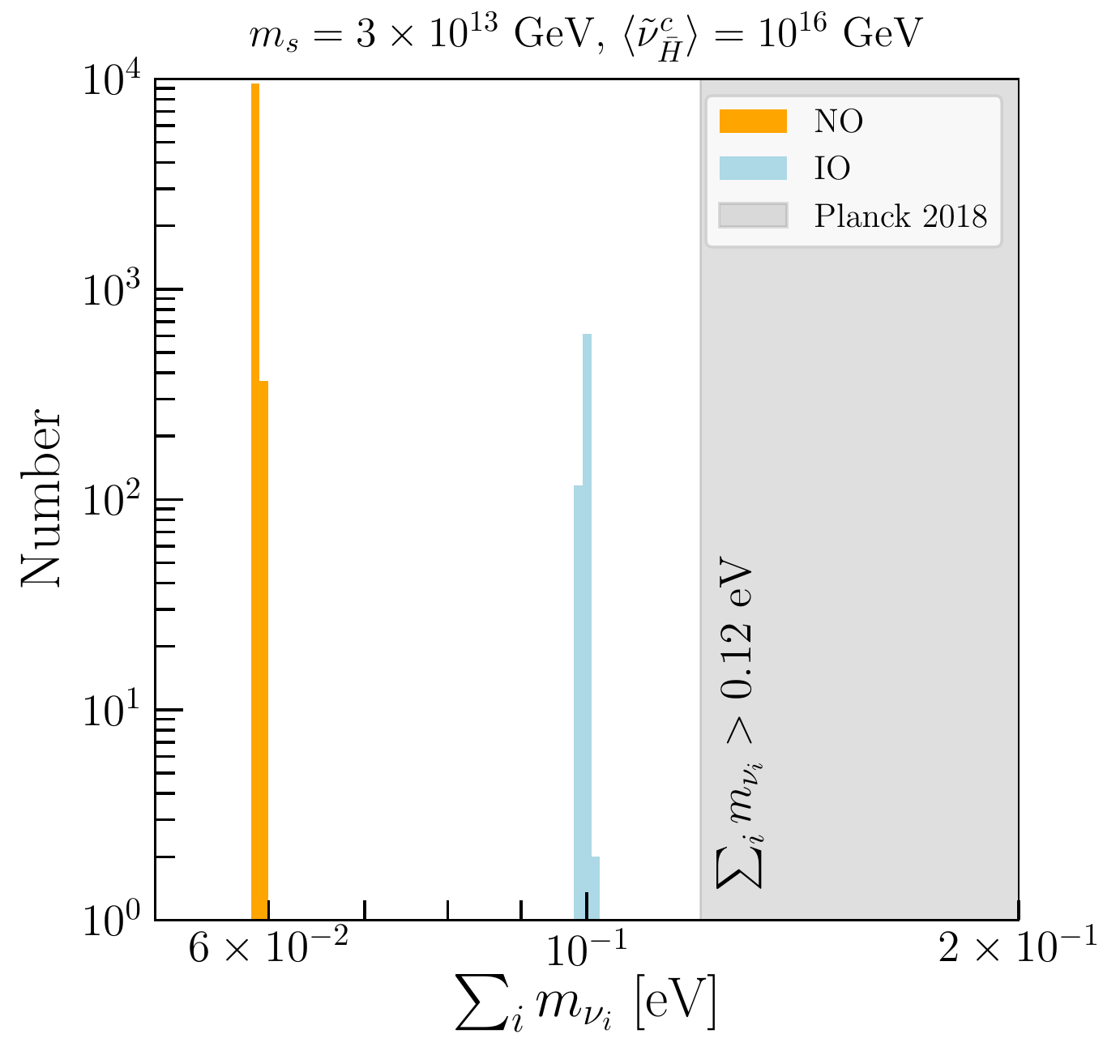}
       \caption{\it Results from our numerical scan over $\lambda_6$ values in the NO and IO scenarios (orange shading and blue dashed, respectively). Left panel: Histogram of values of the lightest neutrino mass.
       Right panel: Histogram of values of the sum of light neutrino masses, showing also the current Planck upper limit~\cite{Aghanim:2018eyx} (shaded grey). }
    \label{fig:mlightest}
\end{figure}

In the left panel in Fig.~\ref{fig:mlightest}, we show the distributions of the lightest neutrino mass for the NO (orange shading) and IO (blue dashed). As we see, in both of the cases the lightest neutrino mass is $\lesssim 10^{-5}$~eV. As a result, in the case of NO, the heavier
neutrinos have masses $\simeq \sqrt{\Delta m_{21}^2} = 8.6 \times
10^{-3}$~eV and $\simeq \sqrt{\Delta m_{31}^2} = 5.0 \times 10^{-2}$~eV, while for the IO case, both of the heavier states have masses $\simeq \sqrt{|\Delta m_{32}^2|} = 5.0 \times 10^{-2}$~eV. The sum of the neutrino masses is then given by $\sum_{i} m_{\nu_i} \simeq 0.06$~eV and 0.1~eV for NO and IO, respectively, as shown in the right panel in Fig.~3. These predicted values are below the current limit imposed by Planck 2018 \cite{Aghanim:2018eyx}, $\sum_i m_{\nu_i} < 0.12$~eV, but can be probed in future CMB experiments such as CMB-S4 \cite{Abazajian:2016yjj}. Moreover, the IO case can be probed
in future neutrino-less double beta decay experiments, whereas testing the NO case in these experiments is challenging \cite{Agostini:2017jim}. 

%%%%%%%%%%%%%%%%%%%%%%%%%%%%%%%%%%%%%%%%%%%%%
%%%%%%%%%%%%%%%%%%%%%%%%%%%%%%%%%%%%%%%%%%%%%%%

\begin{figure}[ht!]
\centering
\subcaptionbox{\label{fig:mu_no}
NO
}
{\includegraphics[width=0.49\textwidth]{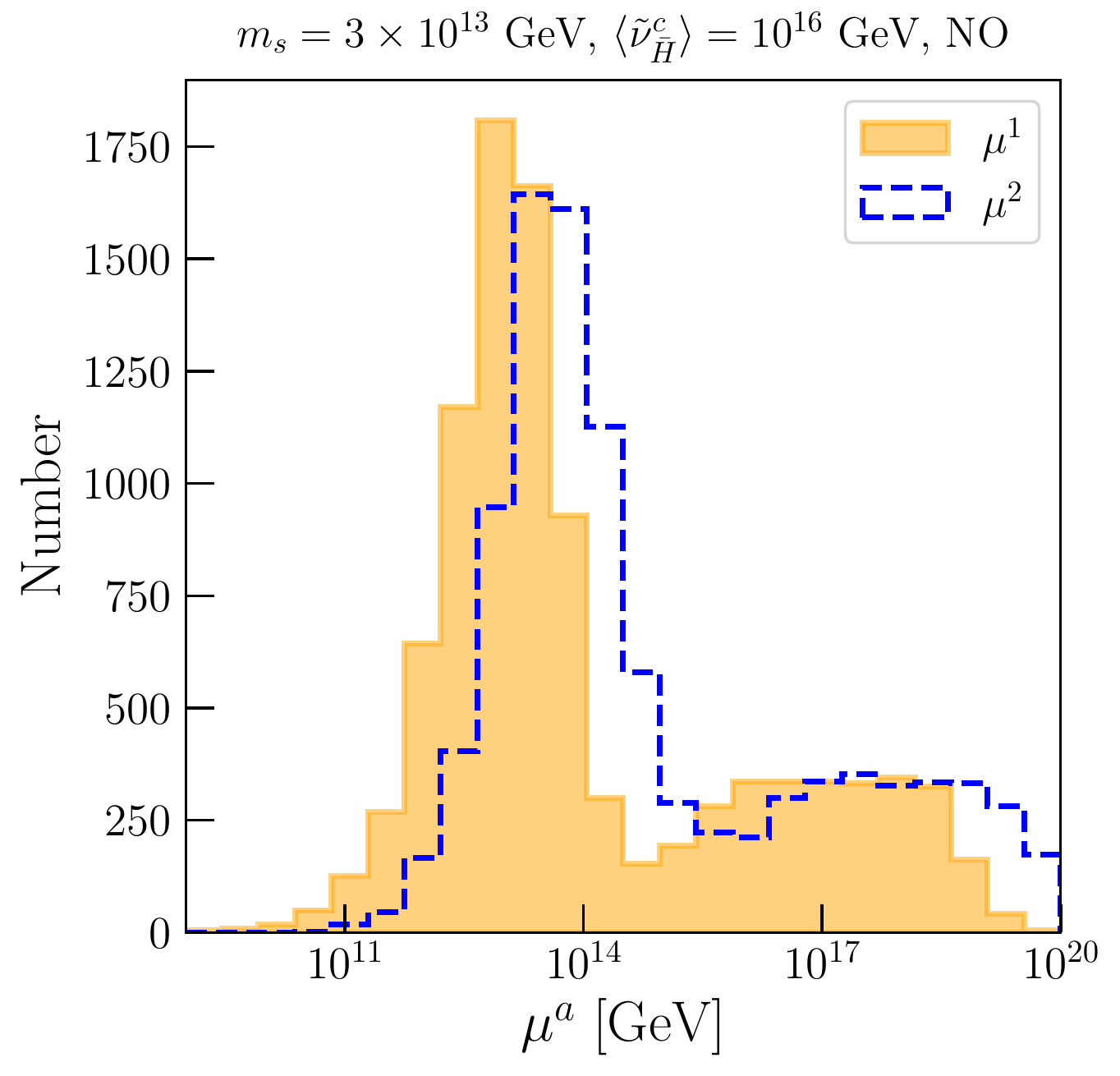}}
\subcaptionbox{\label{fig:mu_io}
IO
}
{\includegraphics[width=0.49\textwidth]{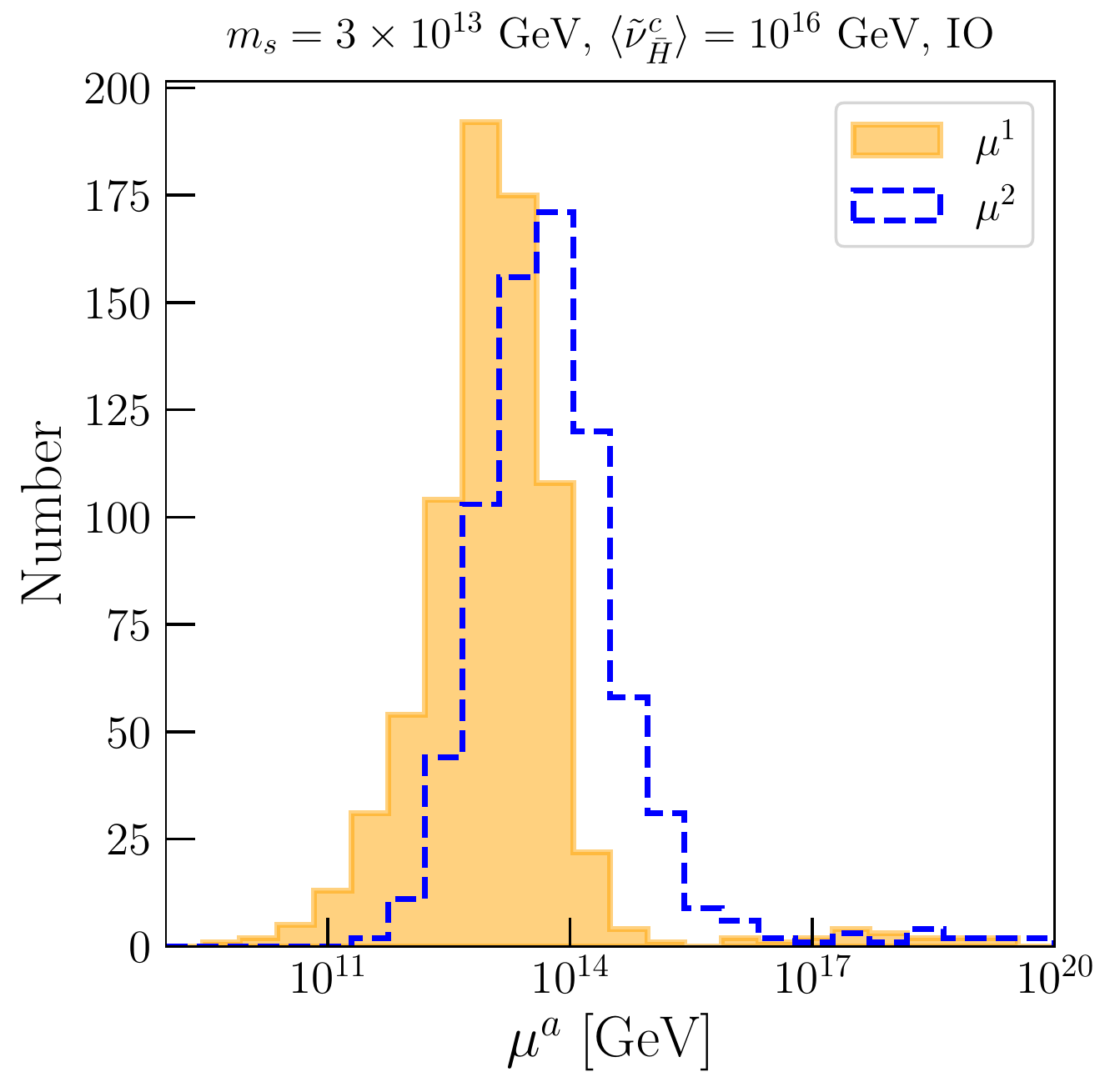}}
\caption{\it  Histograms of the values of $\mu^a$ from our numerical scan over $\lambda_6$ values in the NO and IO scenarios (left and right panels, respectively). 
}
\label{fig:mui}
\end{figure}

In Fig.~\ref{fig:mui} we display histograms of the bilinear couplings $\mu^1$ (orange shading) and $\mu^2$ (blue dashed line) in the NO and IO scenarios (left and right panels, respectively). We see that there is a large range of possible values for these parameters in the NO scenario, peaked around $10^{13}$ to $10^{14}$ GeV, whereas values of $\mu^{1,2}$ in the IO scenario are largely limited to a relatively narrow range around these values. We also find that there is a small number of parameter points that predict $\mu^a > M_P$. Such cases may be disfavored if we consider a plausible ultraviolet completion of our model, specifically in a string theory. In the following analysis, however, we just adopt a bottom-up approach and include these parameter points, while noting that the exclusion of these points do not affect our consequences below. 

%%%%%%%%%%%%%%%%%%%%%%%%%%%%%%%%%%%%%%%%%%%%%%%%%%%%%%%%%%%%%%%%
%%%%%%%%%%%%%%%%%%%%%%%%%%%%%%%%%%%%%%%%%%%%%%%%%%%%%%%%%%%%%%%%

\begin{figure}[ht]
\centering
\subcaptionbox{\label{fig:mnur_no}
NO
}
{\includegraphics[width=0.49\textwidth]{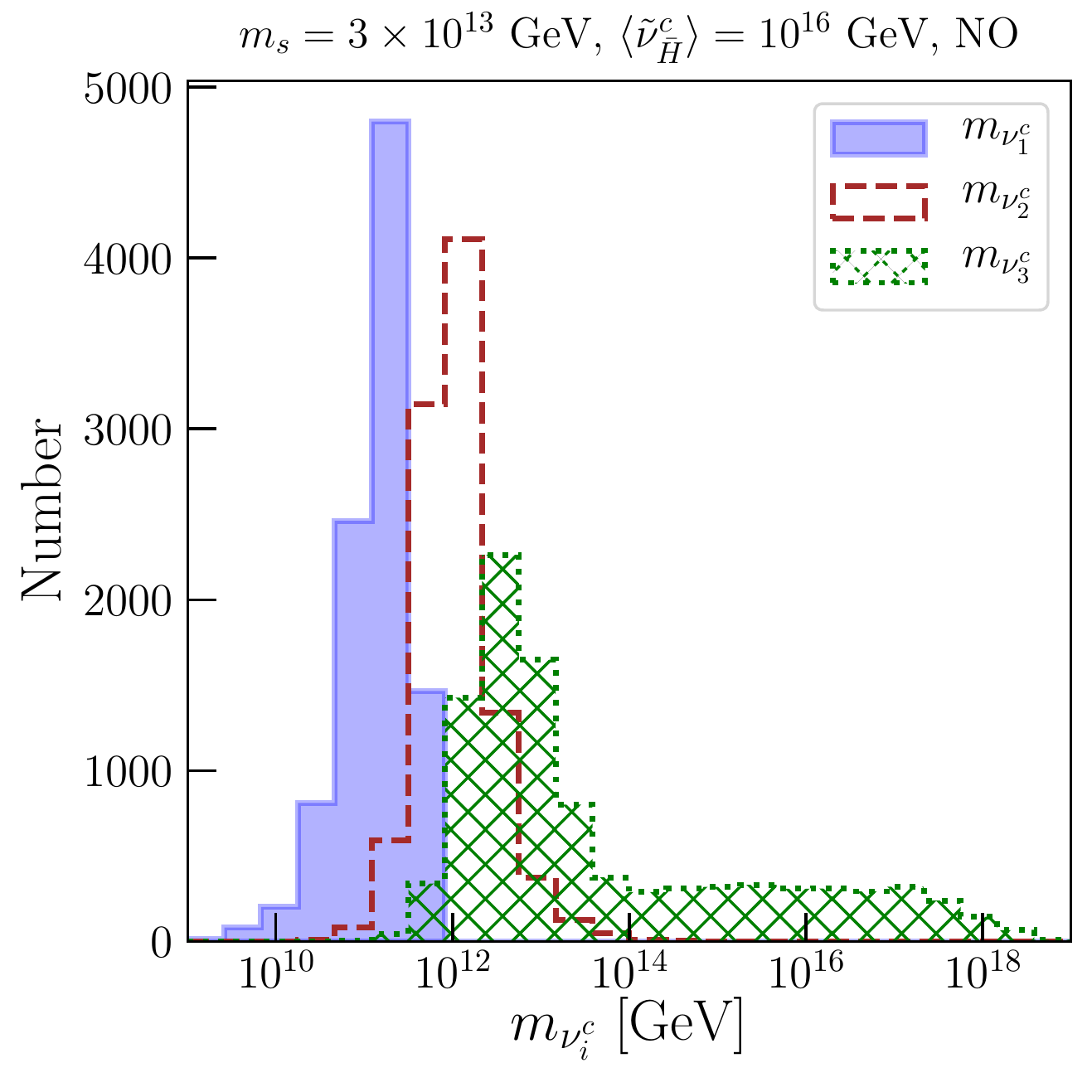}}
\subcaptionbox{\label{fig:mnur_io}
IO
}
{\includegraphics[width=0.49\textwidth]{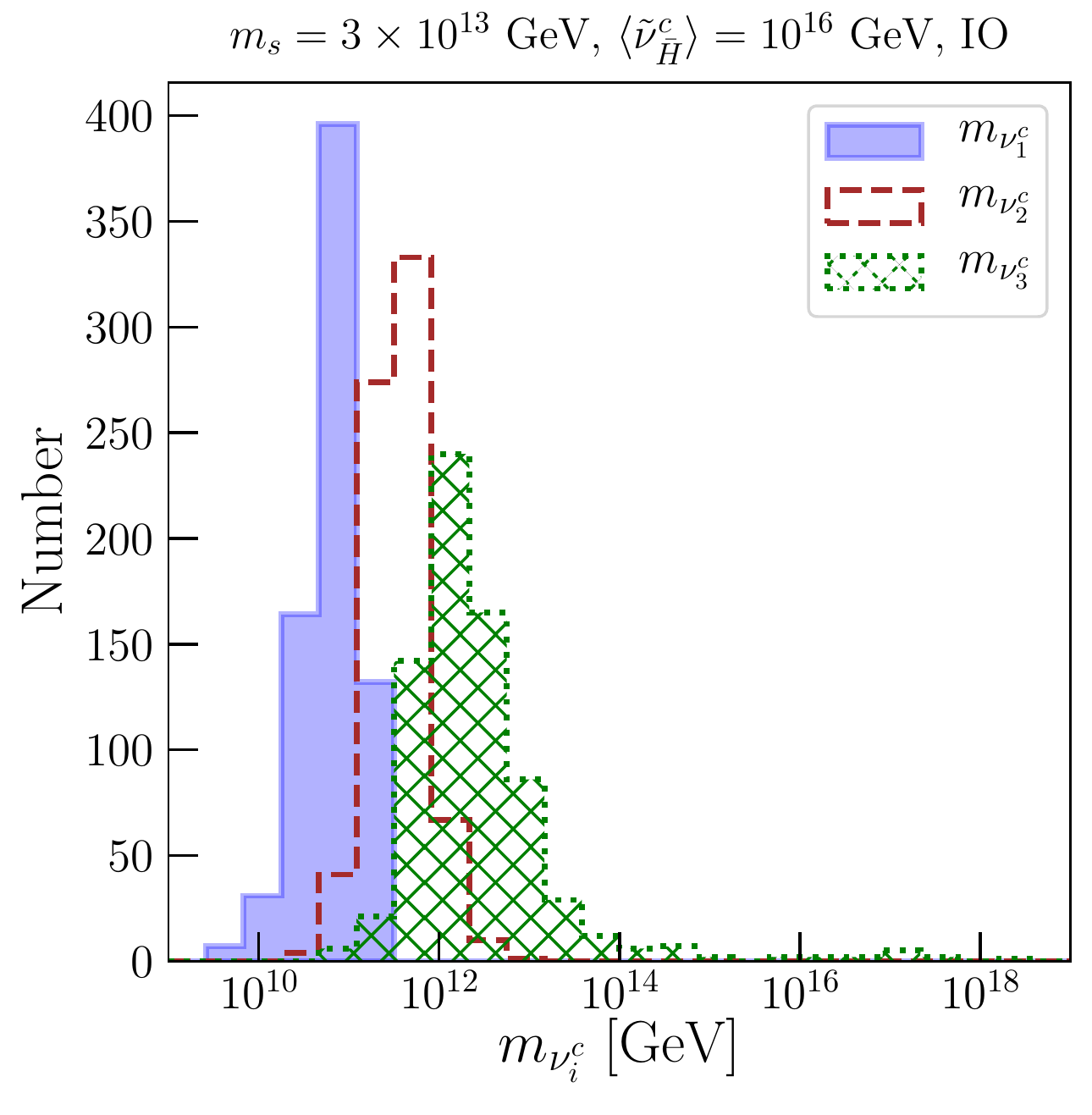}}
\caption{\it Histograms of the spectrum of right-handed neutrino masses from our numerical scan over $\lambda_6$ values in the NO and IO scenarios (left and right panels, respectively). 
}
\label{fig:mnur}
\end{figure}

Fig.~\ref{fig:mnur} shows histograms of the right-handed neutrino masses $m_{\nu^c_1}$ (blue) $m_{\nu^c_2}$ (brown dashed) and $m_{\nu^c_3}$ (green hatching) in the NO and IO scenarios (left and right panels, respectively). In both cases, their distributions are peaked around $10^{11}$, $10^{12}$ and $10^{13}$, respectively. However, possible values of $m_{\nu^c_3}$ in the NO scenario extend to $\sim 10^{18}$~GeV, whereas its values in the IO scenario extend only to $\sim 10^{14}$~GeV. In addition, for most of the parameter points, the right-handed neutrino masses are larger than the critical temperature of the GUT phase transition, $T_c \simeq 0.47 \Lambda_c \gtrsim 10^{10}$~GeV \cite{egnno3}, which justifies the assumption made in the previous Section~\footnote{For a small number of parameter points, the mass of the lightest right-handed neutrino may be less than the critical temperature $T_c$. In this case, we need to take account of the washout effect due to the inverse decay process of $\nu^c_1$ and $\tilde{\nu}^c_1$ in the calculation of the lepton asymmetry. }.

%%%%%%%%%%%%%%%%%%%%%%%%%%%%%%%%%%%%%%%%%%%%%%%%%%%%%%%%%%%%%%%%
%%%%%%%%%%%%%%%%%%%%%%%%%%%%%%%%%%%%%%%%%%%%%%%%%%%%%%%%%%%%%%%%

\begin{figure}[!ht]
\centering
{\includegraphics[width=0.49\textwidth]{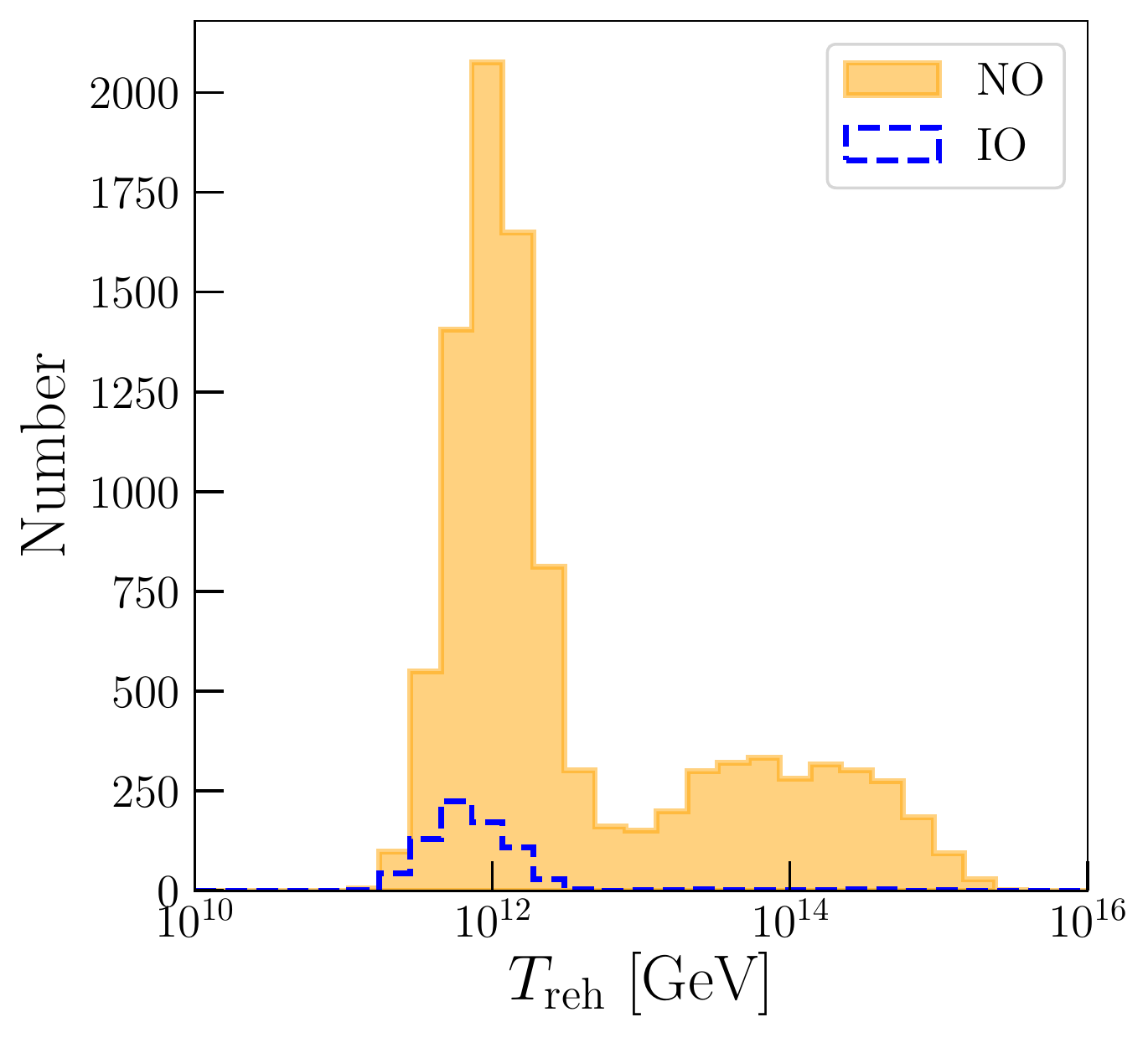}} 
\caption{\it
Histograms of the values of $T_{\rm reh}$ that result from the numerical scan of $\lambda_6$ for the NO and IO cases (orange shading and blue dashed, respectively).
}
\label{fig:tr}
\end{figure}

Histograms of the reheating temperature $T_{\rm reh}$ in the NO and IO scenarios (orange shading and blue dashed, respectively) are shown in Fig.~\ref{fig:tr}. We see that values of $T_{\rm reh} \sim 10^{12}$~GeV are favoured, though much larger values $\lesssim 10^{15}$~GeV are possible in the NO scenario. 
In any case, all of the parameter points predict $T_{\rm reh} \gg \overline{M} \equiv (m_\Phi M_{\rm GUT})^{1/2} \simeq 10^{10}$~GeV, and therefore the strong reheating condition is satisfied as long as $\Lambda_c \gtrsim 2.3 \overline{M}$ \cite{egnno3}.
We note here that in the NO case the range of favoured reheating temperatures includes those for which $T_{\rm reh}\gtrsim m_{s}$. When this is the case the simple picture of perturbative reheating that we have used fails, and thermal and non-thermal in-medium effects, and/or non-perturbative particle production can become relevant and significantly alter the relaxation rate of the inflaton~\cite{reheat}. We nevertheless formally identify $T_{\rm reh}$ as the function of $\lambda_6$ given by (\ref{eq:treh}), keeping in mind that the physical temperature of the plasma after the complete decay of the inflaton may be different. We leave the careful exploration of this ``very strong'' reheating scenario for the future.

%%%%%%%%%%%%%%%%%%%%%%%%%%%%%%%%%%%%%%%%%%%%%%%%%%%%%%%%%%%%%%%%%%%%
%%%%%%%%%%%%%%%%%%%%%%%%%%%%%%%%%%%%%%%%%%%%%%%%%%%%%%%%%%%%%%%%%%%%%

\begin{figure}[ht!]
\centering
{\includegraphics[width=0.49\textwidth]{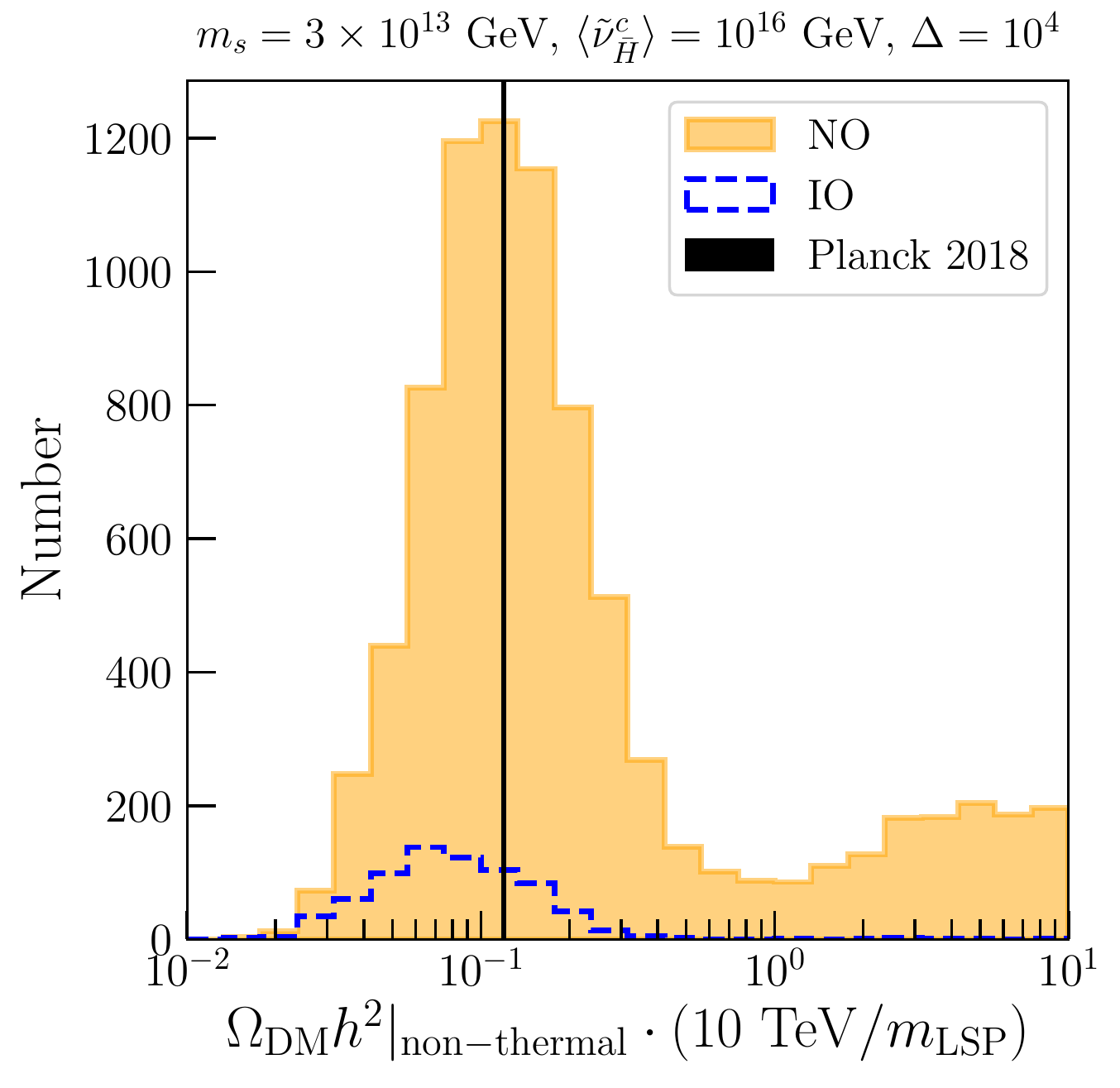}} 
\caption{\it
Histograms of the values of $\Omega_{\rm DM} h^2$ found in the numerical scan of $\lambda_6$ for the NO and IO cases (orange shading and blue dashed, respectively) with $\Delta = 10^4$. The black band shows the Planck 2018 value of the dark matter density: $\Omega_{\rm DM}h^2 = 0.120(1)$ \cite{Aghanim:2018eyx}. 
}
\label{fig:omdm}
\end{figure}

We show in Fig.~\ref{fig:omdm} the distribution of the non-thermal dark matter density produced by gravitino decays in the solutions for $\lambda_6$, where we set the entropy dilution factor to be $\Delta = 10^4$. We find that many parameter solutions predict $\Omega_{\rm DM} h^2 \simeq 10^{-1}$, in good agreement with the dark matter density observed by the Planck collaboration, $\Omega_{\rm DM}h^2 = 0.120(1)$ \cite{Aghanim:2018eyx} (shown as the black band in Fig.~\ref{fig:omdm}), for $m_{\rm LSP} = 10$~TeV, corresponding to $T_{\rm reh} \simeq 10^{12}$~GeV (see Eq.~\eqref{eq:odm}). We also find that some solutions overproduce dark matter for $m_{\rm LSP} = 10$~TeV, as large as $\Omega_{\rm DM} h^2 \simeq 10$, corresponding to
a reheating temperature as high as $T_{\rm reh} \simeq 10^{14}$~GeV. In this case, we need a smaller $m_{\rm LSP}$ and/or a larger $\Delta$ to make these solutions consistent with the observation. 

%%%%%%%%%%%%%%%%%%%%%%%%%%%%%%%%%%%%%%%%%%%%%%%%%%%%%%%%%%%%%%%%%%%%
%%%%%%%%%%%%%%%%%%%%%%%%%%%%%%%%%%%%%%%%%%%%%%%%%%%%%%%%%%%%%%%%%%%%%

The late-time decay of gravitinos is potentially dangerous since it may spoil the successful predictions of BBN. For our default value of the supersymmetry breaking scale, $\simeq 10$~TeV, the lifetime of the gravitino is $\lesssim 100$~s, for which  stringent bounds are given by D/H destruction and $^4$He overproduction \cite{ceflos,Kawasaki:2017bqm}. In
our scenario, for most of the parameter points, the gravitino abundance is sufficiently reduced by the late-time entropy production that the BBN bound is evaded for $m_{3/2} = 10$~TeV. As seen in Figs.~\ref{fig:tr} and \ref{fig:omdm}, however, there are some parameter regions where the reheating temperature is as high as $\gtrsim 10^{14}$~GeV. In this case, as noted above, the relic density would be excessive unless
either the neutralino mass was significantly
reduced, creating a hierarchy between the neutralino
and gravitino masses, or the entropy release $\Delta > 10^4$, in which case the constraint from BBN is also relaxed.

\begin{figure}[!ht]
\centering
\subcaptionbox{\label{fig:baup}
$n_B/s > 0$
}
{\includegraphics[width=0.49\textwidth]{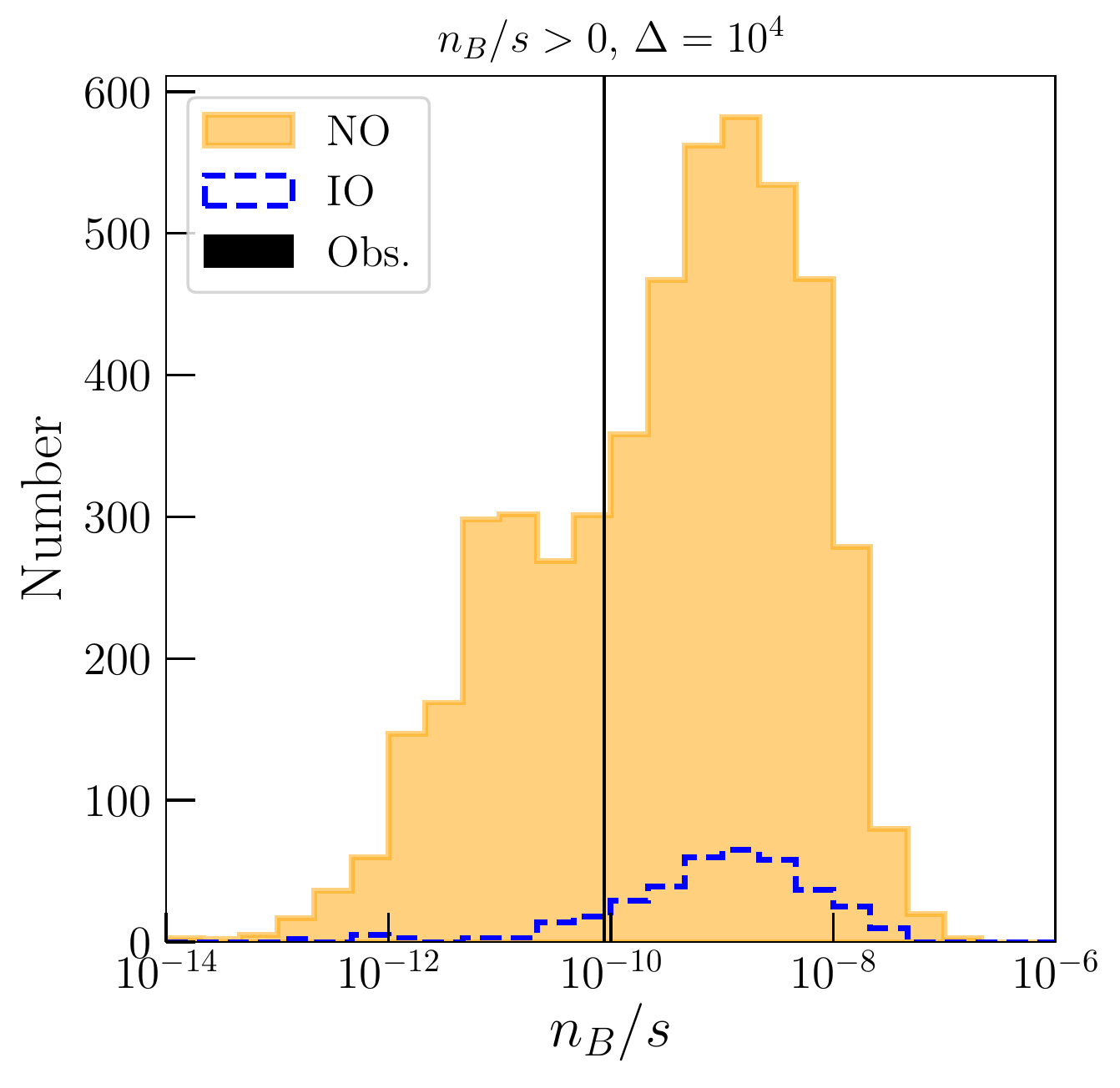}}
\subcaptionbox{\label{fig:baum}
$n_B/s < 0$
}
{\includegraphics[width=0.49\textwidth]{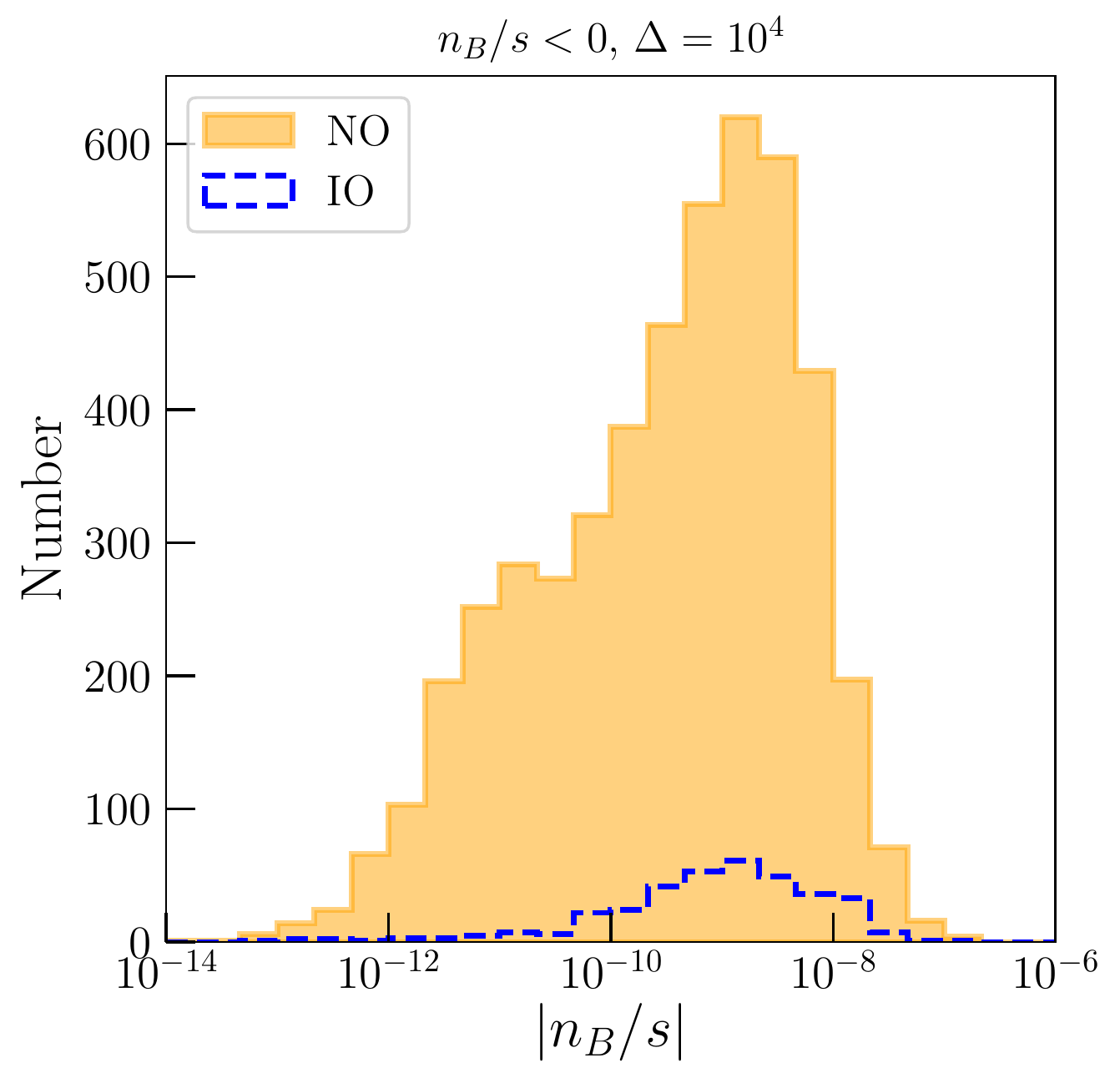}}
\caption{\it Histograms of values of $n_B/s$ in the NO and IO scenarios (orange shading and blue dashed, respectively) for $n_B/s > 0$ (left) and $<0$ (right), assuming an entropy factor $\Delta = 10^4$. The vertical black solid line shows the observed value. 
}
\label{fig:bauhist}
\end{figure}

\begin{figure}[ht]
\centering
    \includegraphics[width=0.49\textwidth]{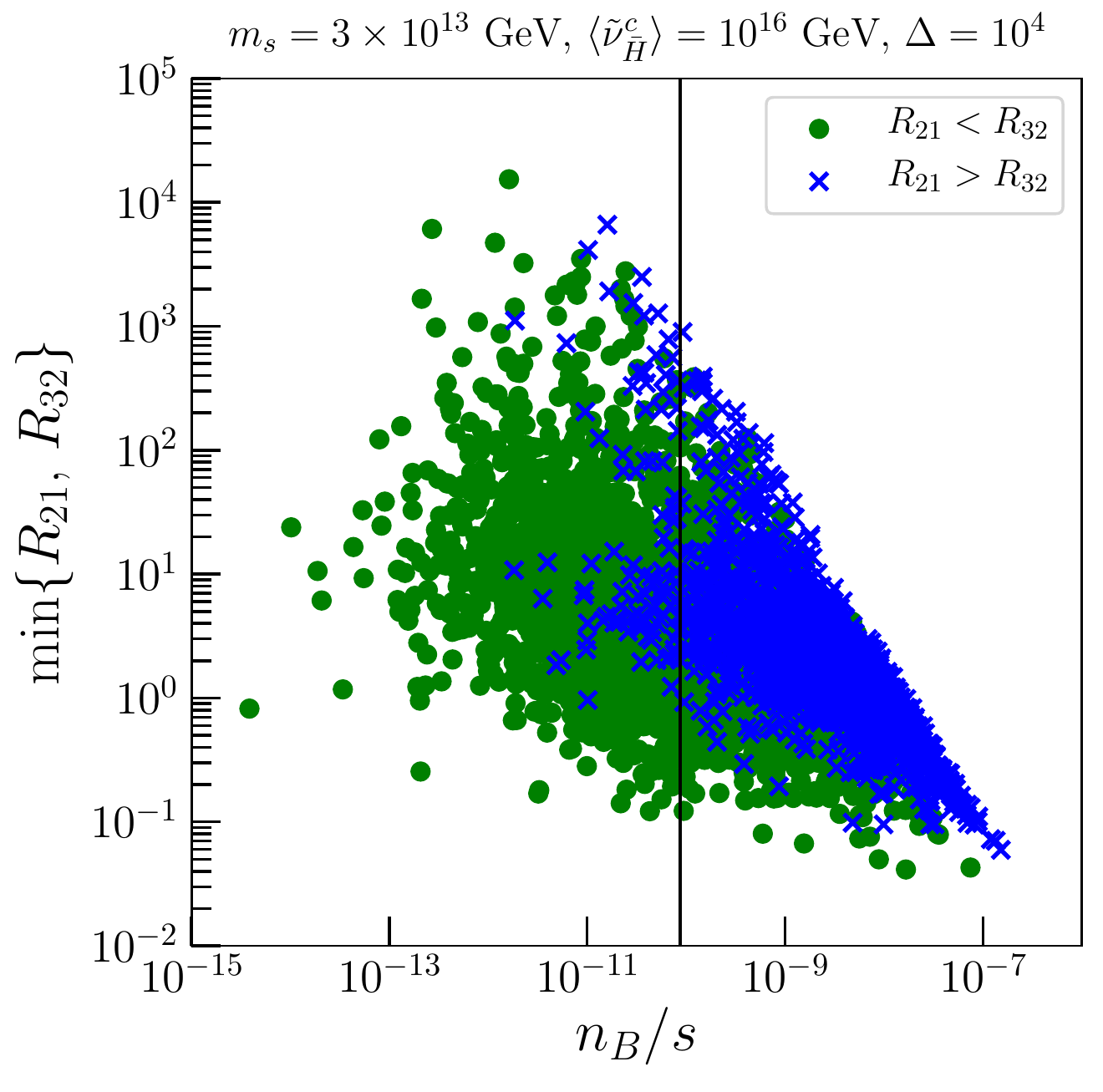}
    \caption{\it Scatter plot of the baryon asymmetry vs the degeneracy of right-handed neutrino masses.
    }
    \label{fig:bau_rat}
\end{figure}

In Fig.~\ref{fig:bauhist}, we show histograms of $n_B/s$ assuming an entropy factor $\Delta = 10^4$; $n_B/s > 0$ in Fig.~\ref{fig:baup} and $n_B/s < 0$ in Fig.~\ref{fig:baum}. As we see, both positive and negative baryon asymmetries can be obtained in both the NO and IO scenarios, without any preference. 
In particular, the observed value (in both magnitude and sign) of the baryon asymmetry, namely
$n_B/s = 0.87 \times 10^{-10}$ \cite{Aghanim:2018eyx}, shown as
the vertical solid line in Fig.~\ref{fig:baup}, can easily be explained in our scenario. It is apparent that a value of $\Delta$ much more than two orders of magnitude larger would be unlikely to yield an acceptable value of $n_B/s$.
We also note that the predicted value of $|n_B/s|$ can be larger than that estimated in Refs.~\cite{egnno2, egnno3}. To see this reason, notice that the mass function $g(x)$ in Eq.~\eqref{eq:gx} gets significantly increased in the limit $x \to 1$:
\begin{equation}
    g (x) \to - \frac{2}{x-1} -1 - \ln(2) \qquad (x\to 1)~.
\end{equation}
This enhancement occurs when there is mass degeneracy in the right-handed neutrino mass spectrum, i.e., $m_{\nu^c_1} \simeq m_{\nu^c_2}$ or $m_{\nu^c_2} \simeq m_{\nu^c_3}$. In Fig.~\ref{fig:bau_rat}, we plot $\text{min}\{R_{21}, R_{31}\}$ against $n_B/s$, where 
\begin{equation}
    R_{ij} \equiv \frac{m_{\nu^c_i} - m_{\nu^c_j}}{m_{\nu^c_j}} 
\end{equation}
quantifies the degree of mass degeneracy. The green dots (blue crosses) correspond to the case $R_{21} < R_{32}$ ($R_{21} > R_{32}$). As we see, $n_B/s \sim 10^{-7}$ can be obtained only when at least a pair of right-handed neutrinos are degenerate in mass at $\lesssim {\cal O} (10)$\% level, a possibility that was not considered in Refs~\cite{egnno2, egnno3}. For a smaller value of $n_B/s$, we do not need such mass degeneracy. In particular, $n_B/s \simeq 0.87 \times 10^{-10}$ can be obtained even for a hierarchical right-handed neutrino mass spectrum, which is consistent with the estimation done in the previous papers \cite{egnno2, egnno3}.

%%%%%%%%%%%%%%%%%%%%%%%%%%%%%%%%%%%%%%%%%%%%%%%%%%%%%%%%%%%%%%%%
%%%%%%%%%%%%%%%%%%%%%%%%%%%%%%%%%%%%%%%%%%%%%%%%%%%%%%%%%%%%%%%%

%%%%%%%%%%%%%%%%%%%%%%%%%%%%%%%%%%%%%%%%%%%%%
%%%%%%%%%%%%%%%%%%%%%%%%%%%%%%%%%%%%%%%%%%%%%%%
%%%%%%%%%%%%%%%%%%%%%%%%%%%%%%%%%%%%%%%%%%%%%%%%%%%

\begin{figure}[!ht]
\centering
{\includegraphics[width=0.49\textwidth]{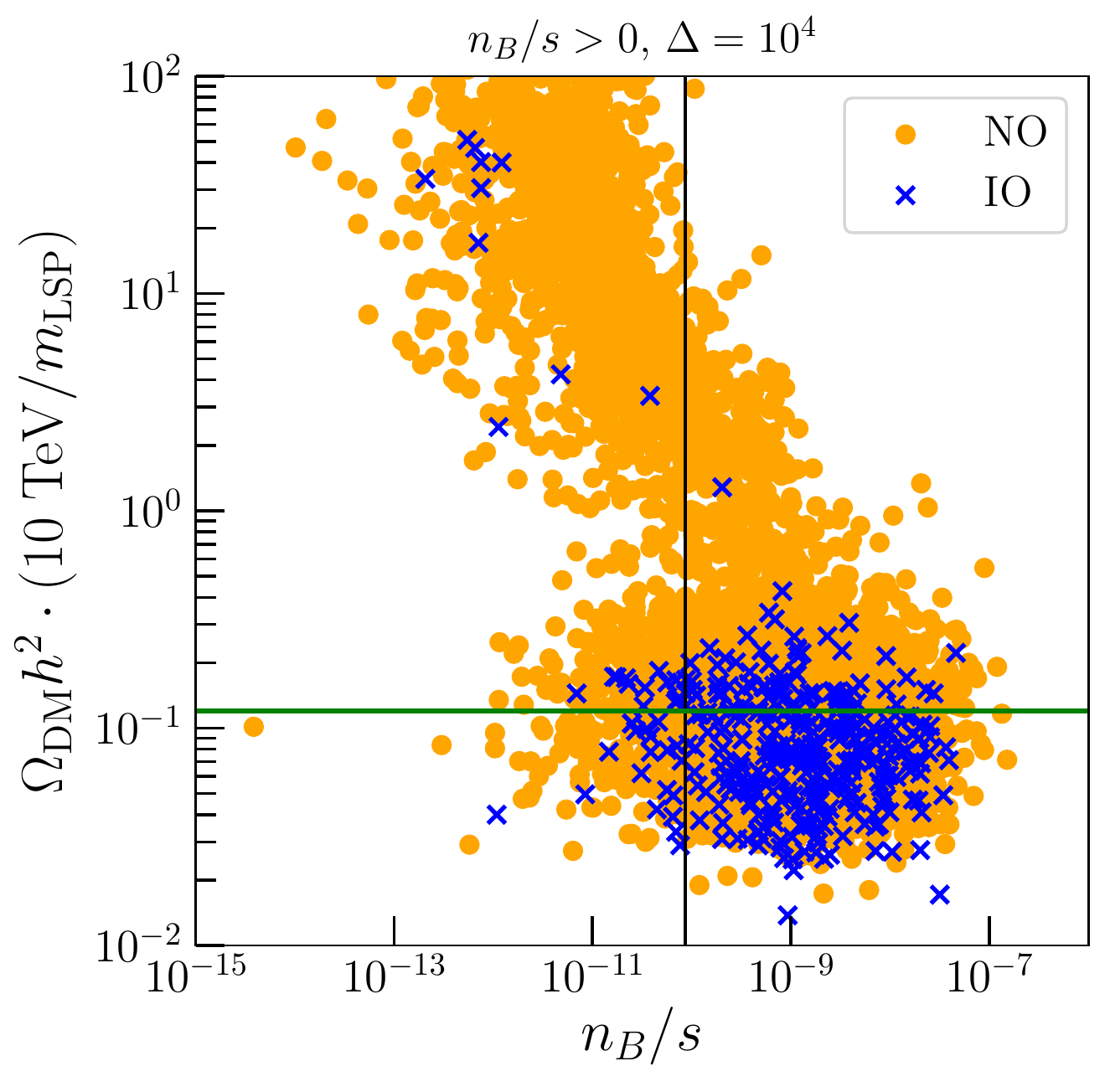}} 
\caption{\it 
Scatter plot of $n_B/s$ vs the non-thermal contribution to the LSP abundance, assuming $m_{\rm LSP} = 10$~TeV and $\Delta = 10^4$, with the
 observed values shown as the horizontal green and vertical black lines,
 respectively. 
}
\label{fig:bau_omdm}
\end{figure}

In Fig.~\ref{fig:bau_omdm}, we 
%have also evaluated 
plot the non-thermal contribution to the LSP abundance from gravitino
decay, assuming $m_{\rm LSP} = 10$~TeV,
%, and plot its value 
against the baryon asymmetry predicted at the same parameter point,
assuming $\Delta = 10^4$. The
vertical black and horizontal green lines show, respectively, the observed values of
baryon asymmetry and dark matter abundance $\Omega_{\rm DM} h^2 = 0.12$
\cite{Aghanim:2018eyx}.
We find that most of the points
predict $n_B/s \lesssim {\cal O}(10^{-7})$ and $\Omega_{\rm DM} h^2 \gtrsim
{\cal O}(10^{-2})$. 
In particular, there are many solutions where $n_B/s \simeq 0.87 \times 10^{-10}$ and the non-thermal component
of the LSP abundance from gravitino decays accounts for the entire dark matter density $\Omega_{\rm DM} h^2 \simeq 0.12$. For such parameter points, one must ensure that the thermal relic of the LSP is sufficiently depleted, which is obtained easily if $\Delta \sim 10^4$, as we see in the next Section.

To illustrate that a good solution can be found for a generic choice of the $\lambda_6$ matrix, in Table~\ref{tab:example}, we show the predicted values of physical observables for 
\begin{equation}
    \lambda_6 =
\begin{pmatrix}
-4.5 \times 10^{-5} - 2.7 \times 10^{-4} i
& 6.4 \times 10^{-4} + 3.9 \times 10^{-4} i
& 5.0 \times 10^{-4} - 4.4 \times 10^{-5} i \\
8.2 \times 10^{-5} -3.2 \times 10^{-4} i 
& -4.9 \times 10^{-4} + 3.4 \times 10^{-4} i
& 6.4 \times 10^{-4} + 1.4 \times 10^{-4} i \\
-3.3 \times 10^{-4} + 3.1 \times 10^{-4} i 
& 1.0 \times 10^{-5} - 2.2 \times 10^{-4} i
& -1.2 \times 10^{-3} + 4.2 \times 10^{-4}i
\end{pmatrix}
~.
\label{eq:lambda6example}
\end{equation}
In general, we find that a generic choice of $\lambda_6$ with absolute values of ${\cal O}(10^{-4})$ can reproduce the observed baryon asymmetry and the dark matter density. In this case, the right-handed neutrino mass spectrum is moderately hierarchical, and all of the singlet fermions $\tilde{\phi}_a$ have masses much larger than the right-handed neutrinos. We add also that for the $\lambda_6$ matrix (\ref{eq:lambda6example}) and $\Delta \sim 10^4$ the calculations in~\cite{egnno3} indicate that $n_s \simeq 0.961$, within the 68\% CL range $n_s = 0.9645 \pm 0.0042$ allowed by Planck and other data~\cite{Aghanim:2018eyx}.

%%%%%%%%%%%%%%%%%
%%%%%%%%%%%%%%%%%%%%%%%%%%%%%%%%%%%%%%%%%%%%%%%%%%%%%%%%%%%%%%%%%%%%%

%%%%%%%%%%%%%%%%%%%%%%%%%%%%%%%%%%%%%%%%%%%%%%%%%%%%%%%%%%%%%
\begin{table}[!ht]
\begin{center}
\begin{tabular}{l|c|| l|c} 
\hline
\hline
$m_{1}$ [eV] & $1.1 \times 10^{-7}$ &
$m_{\nu^c_1}$ [GeV] & $3.8\times 10^{10}$ \\ 
$m_{2}$ [eV] & $8.6 \times 10^{-3}$ &
$m_{\nu^c_2}$ [GeV] & $3.3\times 10^{11}$ \\ 
$m_{3}$ [eV] & $5.0 \times 10^{-2}$ &
$m_{\nu^c_3}$ [GeV] & $2.2\times 10^{12}$ \\ 
$\sum_i m_i$ [eV] & $5.9 \times 10^{-2}$&
$T_{\rm reh}$ [GeV] & $1.1\times 10^{12}$ \\ 
$|\mu_1|$ [GeV] & $1.2\times 10^{14}$& 
$\Omega_{\rm DM} h^2|_{\rm non-thermal}$ & $0.12$  \\
$|\mu_2|$ [GeV] & $1.4\times 10^{15}$ &
$n_B/s$ &$0.88\times 10^{-10}$ \\
\hline
\hline
\end{tabular}
\caption{\it The predicted values of physical observables for the $\lambda_6$ coupling matrix given in Eq.~\eqref{eq:lambda6example} with $m_s = 3\times 10^{13}$~GeV, $\langle \tilde{\nu}_{\bar{H}}^c \rangle = 10^{16}$~GeV,  $m_{\rm LSP} = 10$~TeV, and $\Delta = 10^4$. } 
\label{tab:example}
\end{center}
\end{table}
%%%%%%%%%%%%%%%%%%%%%%%%%%%%%%%%%%%%%%%%%%%%%%%%%%%%%%%%%%%%%%

%%%%%%%%%%%%%%%%%%%%%%%%%%%%%%%%%%%%%%%%%%%%%
\begin{figure}[!ht]
\centering
    \includegraphics[width=0.49\textwidth]{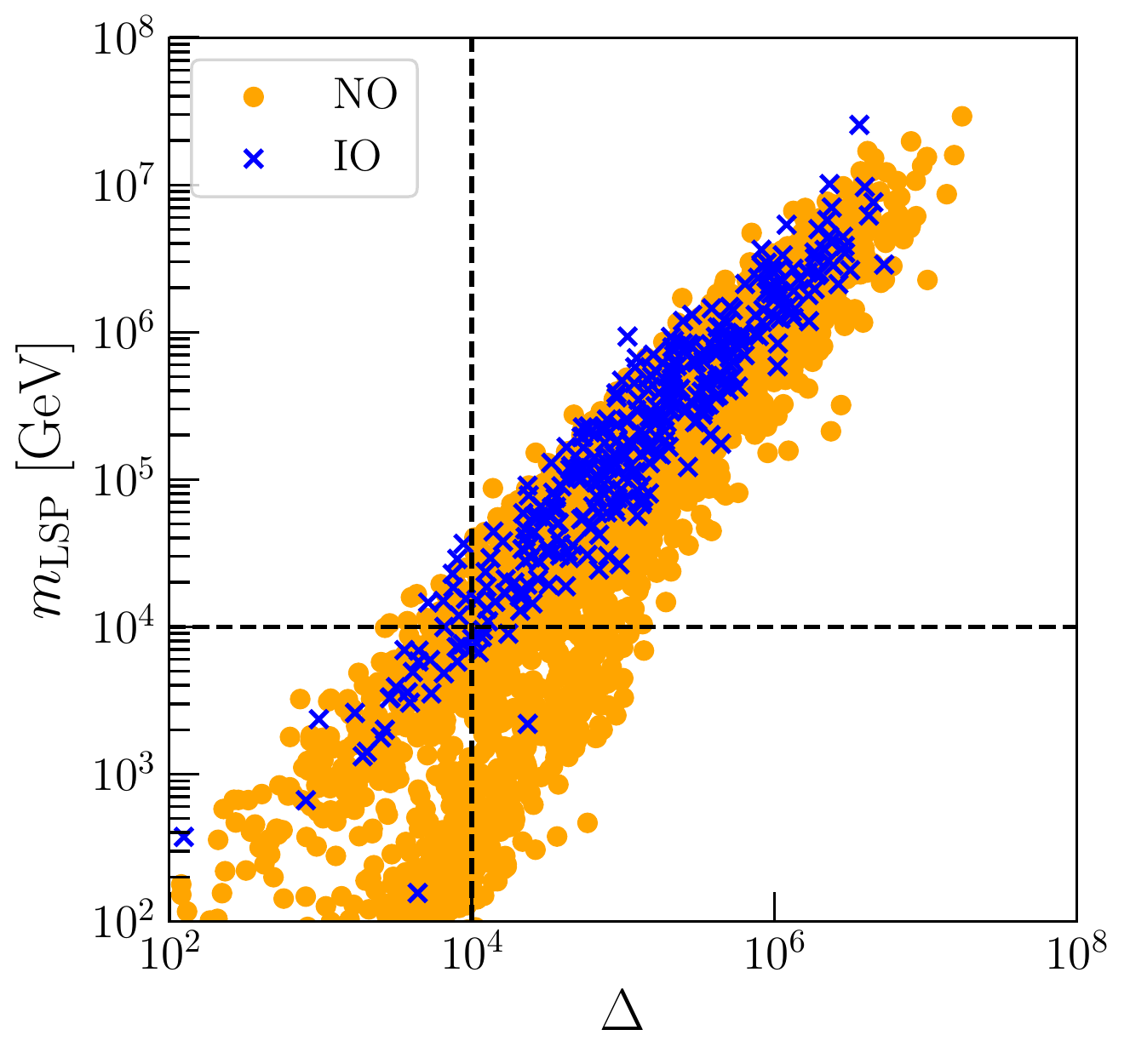}
    \caption{\it 
Scatter plot of the entropy factor $\Delta$ vs $m_{\rm LSP}$, assuming that all the dark matter is produced non-thermally and that the measured value of $n_B/s$ is reproduced. The values $\Delta = 10^4$ and $m_{\rm LSP} = 10^4$~GeV that we have used as defaults in previous figures are shown as vertical and horizontal dashed lines.
    }
    \label{fig:delta_mlsp}
\end{figure}
%%%%%%%%%%%%%%%%%%%%%%%%%%%%%%%%%%%%%%%%%%%%%%%

Figs.~\ref{fig:omdm} and \ref{fig:bauhist} suggested a preferred range of the entropy factor $\Delta \sim 10^4$. We show in Fig.~\ref{fig:delta_mlsp} a scatter plot of $\Delta$ vs $m_{\rm LSP}$, assuming that all the dark matter is produced non-thermally and requiring that the measured value of $n_B/s$ be reproduced. We see that the values of $\Delta$ range from $\sim 10^2$ to $\sim 10^7$ and that $m_{\rm LSP}$ lies between $\sim 10^2$ and $10^7$~GeV in the NO scenario, with a strong correlation between the two quantities. The vertical and horizontal dashed lines indicate the values $\Delta = 10^4$ and $m_{\rm LSP} = 10^4$~GeV that we have used as defaults in previous figures. In the IO scenario $m_{\rm LSP} \gtrsim 10^3$~GeV, while lower values of $m_{\rm LSP}$ are possible in the NO scenario. However, even in this scenario our default choices $\Delta = 10^4$ and $m_{\rm LSP} = 10^4\,{\rm GeV}$ are quite representative. Therefore, in the following Section we consider some phenomenological aspects of such a heavy sparticle spectrum.

%%%%%%%%%%%%%%%%%%%%%%%%%%%%%%%%%%%%%%%%%%%%%%%%%%%%%%%%%%%%%%%%%%%%%%%%%%%%
\section{Phenomenological Aspects of Entropy Production}
\label{sec:pheno}
%%%%%%%%%%%%%%%%%%%%%%%%%%%%%%%%%%%%%%%%%%%%%%%%%%%%%%%%%%%%%%%%%%%%%%%%%%%%

One of the most striking phenomenological consequences of 
entropy production on the scale discussed above is its effect on 
the low-energy supersymmetric parameter space.
In constrained models such as the CMSSM 
\cite{cmssm,elos,eelnos,eemno,eeloz,ehow++}, 
all soft scalar masses are unified with the same input value 
$m_0$ at some high energy
scale, $M_{in}$, which may be equal to the GUT scale, as in the CMSSM,
or less (as in sub-GUT models~\cite{elos,eelnos,eeloz,subGUT,mcsubgut}), or greater than the GUT scale (as in super-GUT models~\cite{superGUT,emo,emo3,eemno}), and similarly for gaugino masses, $m_{1/2}$, and trilinear terms, $A_0$.
In the absence so far of a positive signal for supersymmetry
at the LHC~\cite{nosusy}, the viability of such constrained models
relies on particular relations between the mass of the LSP
and some other sparticle masses.  
In general, with TeV-scale sparticle masses we expect the relic LSP density left over from thermal freeze-out to be
relatively large, i.e., generically much larger than the cold dark matter
density determined by Planck, $\Omega h^2 \simeq 0.12$ \cite{Aghanim:2018eyx}.

For example, let us consider the representative example of a bino LSP $\chi$
annihilating to SM fermions through sfermion exchange.
Assuming roughly $m_{\tilde f} > m_\chi \gg m_f$,
we can approximate the $p$-wave annihilation cross section~\footnote{The $s$-wave cross section is suppressed relative to the $p$-wave by a factor of $(m_f/m_\chi)^2$.} as \cite{OS}
\beq
\langle \sigma v \rangle \simeq \frac{g_1^4}{32 \pi} \sum_f ({Y_L}_f^4 + {Y_R}_f^4) \frac{m_\chi^2}{m_{\tilde f}^4} x ,
\label{sv}
\eeq
where $g_1$ is the U(1)$_Y$ gauge coupling, $Y_{L,R}$ are the hypercharges
of (left, right)-handed fermions, $m_\chi$ is the bino mass, $m_{\tilde f}$
is a common sfermion mass, and $x = T_f/m_\chi \approx 1/20$ is the
annihilation freeze-out temperature relative to the bino mass. 
The relic density can be approximated by \cite{ehnos,eo}
\beq
\Omega_\chi h^2 \approx 1.9 \times 10^{-11} \left(\frac{T_\chi}{T_\gamma}\right)^3 \sqrt{g_f} \left(\frac{{\rm GeV}^{-2}}{{\frac12 \langle \sigma v \rangle x}}\right) ,
\label{oh2}
\eeq
where the factor $({T_\chi}/{T_\gamma})^3$ accounts for the dilution
of neutralinos from freeze-out to today \cite{oss,ehnos}, and $g_f$ is the 
number of relativistic degrees of freedom at freeze-out. For $m_\chi \sim 100$~GeV and $m_{\tilde f}\sim 350$~GeV, $\Omega_\chi h^2 \sim 0.1$.
However, it is apparent from (\ref{oh2}) that $\Omega_\chi h^2$ scales as $m_{\tilde f}^4/m_\chi^2$, implying that
increasing the supersymmetry-breaking scale by a factor of 100
so that, e.g., $m_\chi \sim 10$ TeV and $m_{\tilde f}\sim 35$ TeV,
we would find $\Omega_\chi h^2$ of order $10^3$.

This argument may be circumvented by relying on coannihilations~\cite{gs}
between the LSP and the next-to-lightest supersymmetric particle (NLSP), such as the
lighter stop~\cite{stopco,eds,eoz,interplay,raza,eeloz,ehow++} or stau~\cite{stau,celmov,delm}. 
For such coannihilations to be
effective at reducing the relic density,
a high degree of degeneracy is needed between the LSP and NLSP masses. Alternatively, the mass of the neutralino may be very close to 1/2 the mass of the heavy Higgs scalar and/or pseudoscalar, leading to rapid $s$-channel annihilations~\cite{funnel}.
Another possibility occurs when $m_0 \gg m_{1/2}$ with 
small (or vanishing) $A_0$. In this regime, the value
of the Higgs mixing parameter $\mu$ is driven to zero, the LSP becomes more
Higgsino-like, and annihilations to electroweak gauge bosons
become significant~\cite{fp}. Again, this requires
a fairly finely-tuned relation between $m_{1/2}$ and $m_0$
for given values of $A_0$ and $\tan \beta$.

Late-time entropy production changes dramatically the 
landscape of allowed models. A factor $\Delta = 10^4$ in entropy production
would imply that the preferred relic density at freeze-out would be increased by $10^4$, corresponding to typical supersymmetry-breaking masses of order 10~TeV as can be inferred from Eqs. (\ref{sv}) and (\ref{oh2}).  To the extent that the model requires $\Delta = 10^4$, one may consider that a supersymmetry-breaking scale of ${\mathcal{O}}(10)$ TeV
is a prediction of the model. Very roughly, we can re-express the relic density
in Eq. (\ref{oh2}) as
\beq
\Omega_\chi h^2 \simeq  10^{-7} {\rm GeV}^{-2} \Delta^{-1} \frac{m_{\tilde f}^4}{m_\chi^2}
\sim 10^3 \Delta^{-1} \left(\frac{m_{\tilde f}}{30 {\rm TeV}}\right)^4 \left(\frac{10 {\rm TeV}}{m_\chi} \right)^2 \, ,
\label{oh22}
\eeq
where the entropy release is given roughly by~\cite{egnno3}
\beq
\Delta \sim 10^4 \left( \frac{30 {\rm TeV}}{m_{\tilde f}} \right)^{1/2} \, ,
\eeq
so that 
\beq
\Omega_\chi h^2 \sim 10^{-1} \left(\frac{m_{\tilde f}}{30 {\rm TeV}}\right)^{9/2} \left(\frac{10 {\rm TeV}}{m_\chi} \right)^2 ,
\label{oh23}
\eeq
where we have assumed that all relevant couplings are of order 1.

Another cosmological consideration that should be taken into account is successful BBN, which requires the  
reheating temperature after the transition to be at least 1~MeV,
so as to ensure a radiation-dominated universe during BBN. The reheating temperature
can be written as \cite{egnno3}
\beq
T_{\rm reh}^\prime \sim 10^{-3} \left(\frac{m_{\tilde f}^3 M_P}{M_{\rm GUT}^2} \right)^{1/2} \sim 1 {\rm MeV} \left( \frac{m_{\tilde f}}{30 {\rm TeV}} \right)^{3/2}\, ,
\label{tr'}
\eeq
and combining Eqs.~\eqref{oh23} and \eqref{tr'}, we can write
\beq
 \Omega_\chi h^2 \sim 0.1 \left( \frac{T_{\rm reh}^\prime}{1 {\rm MeV}} \right)^3 \left(\frac{10 {\rm TeV}}{m_\chi} \right)^2 \, .
\eeq
It is rather remarkable that the late-time reheating temperature in Eq.~(\ref{tr'}) is
just above 1 MeV, as needed to restart 
BBN, for $m_{\tilde f} \gtrsim {\mathcal O}$(10) TeV, while 
the observed value of $\Omega_\chi h^2$ in Eq.~(\ref{oh22}) requires
$m_{\tilde f} \lesssim {\mathcal O}$(10) TeV, thus determining 
supersymmetry breaking scale in our model to be ${\mathcal O}$(10) TeV.
{\it This prediction is consistent with the 
non-observation of SUSY signals 
at LHC so far.}
%Thus while scales much larger than ${\mathcal O}(10)$ TeV, would
%produce a higher reheat temperature, the relic density would be excessive. 

\begin{figure}[htb!]
%\vskip 0.5in
%\vspace*{-0.75in}
%\hspace*{-.70in}
\begin{minipage}{8in}
%\hspace*{-1.2in}
\includegraphics[height=3in]{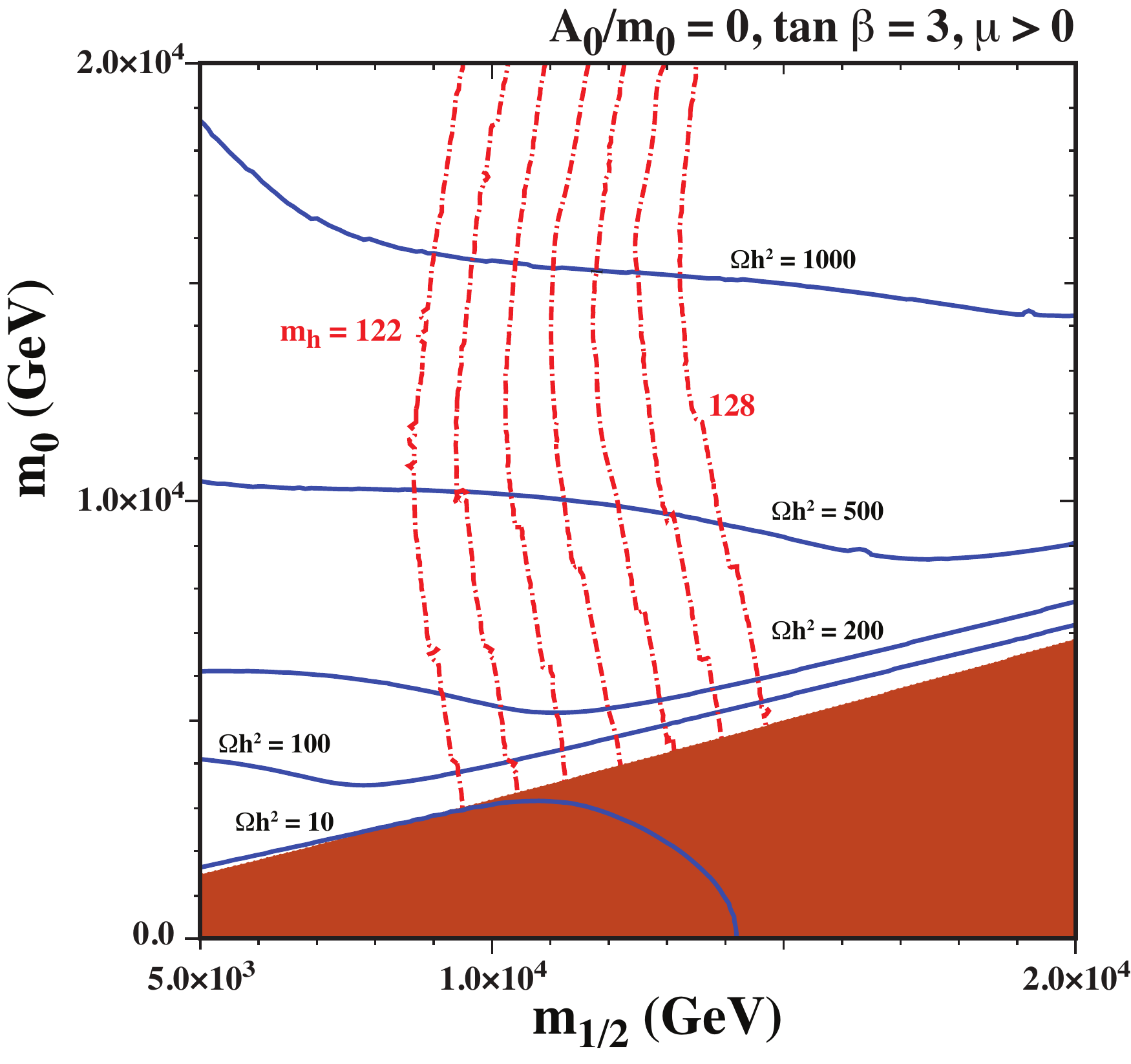}
%\hspace*{-2.7in}
\includegraphics[height=3in]{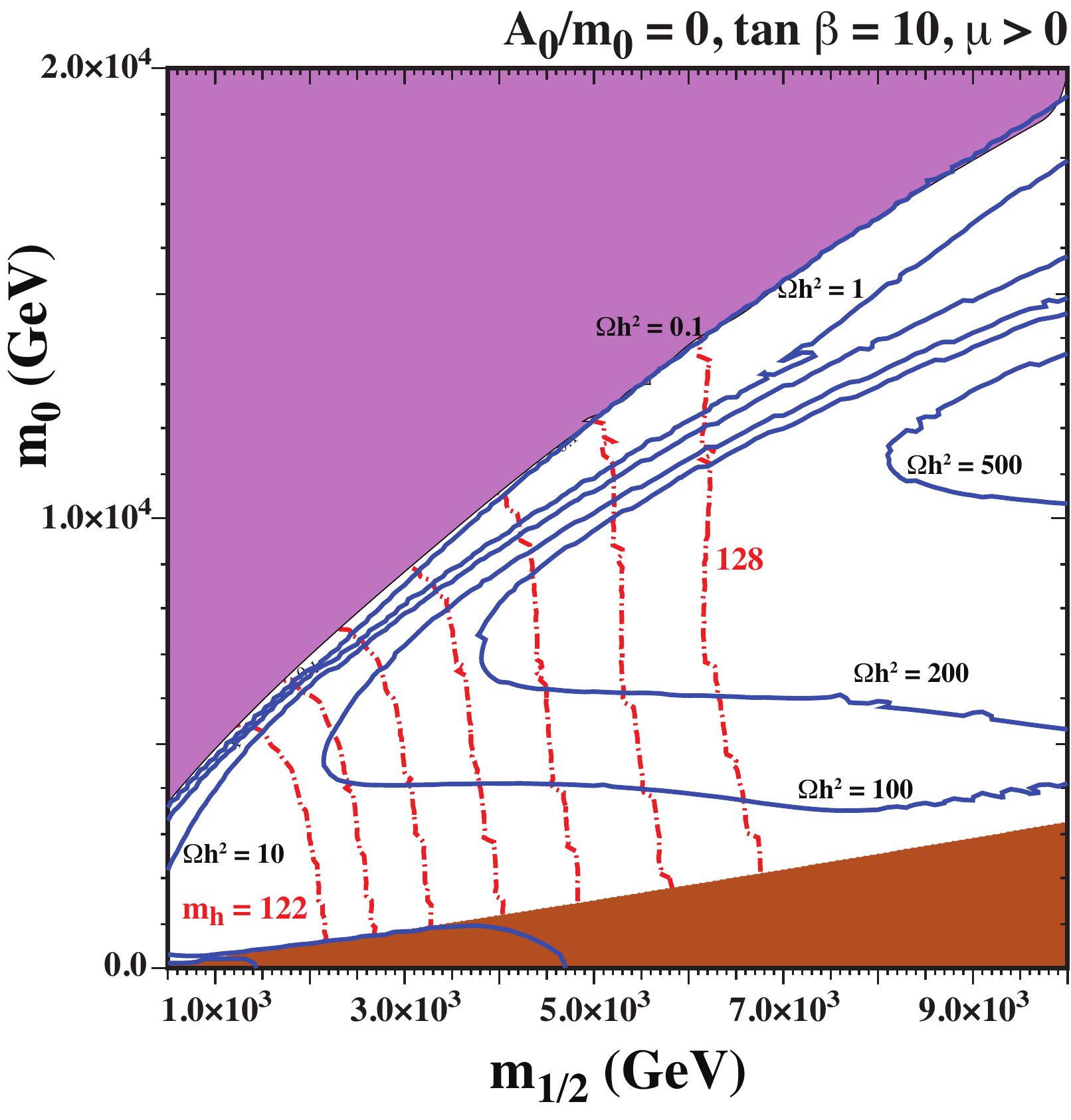}\\
%\vspace*{.5in}
\hfill
%\vspace*{-2in}
\includegraphics[height=3in]{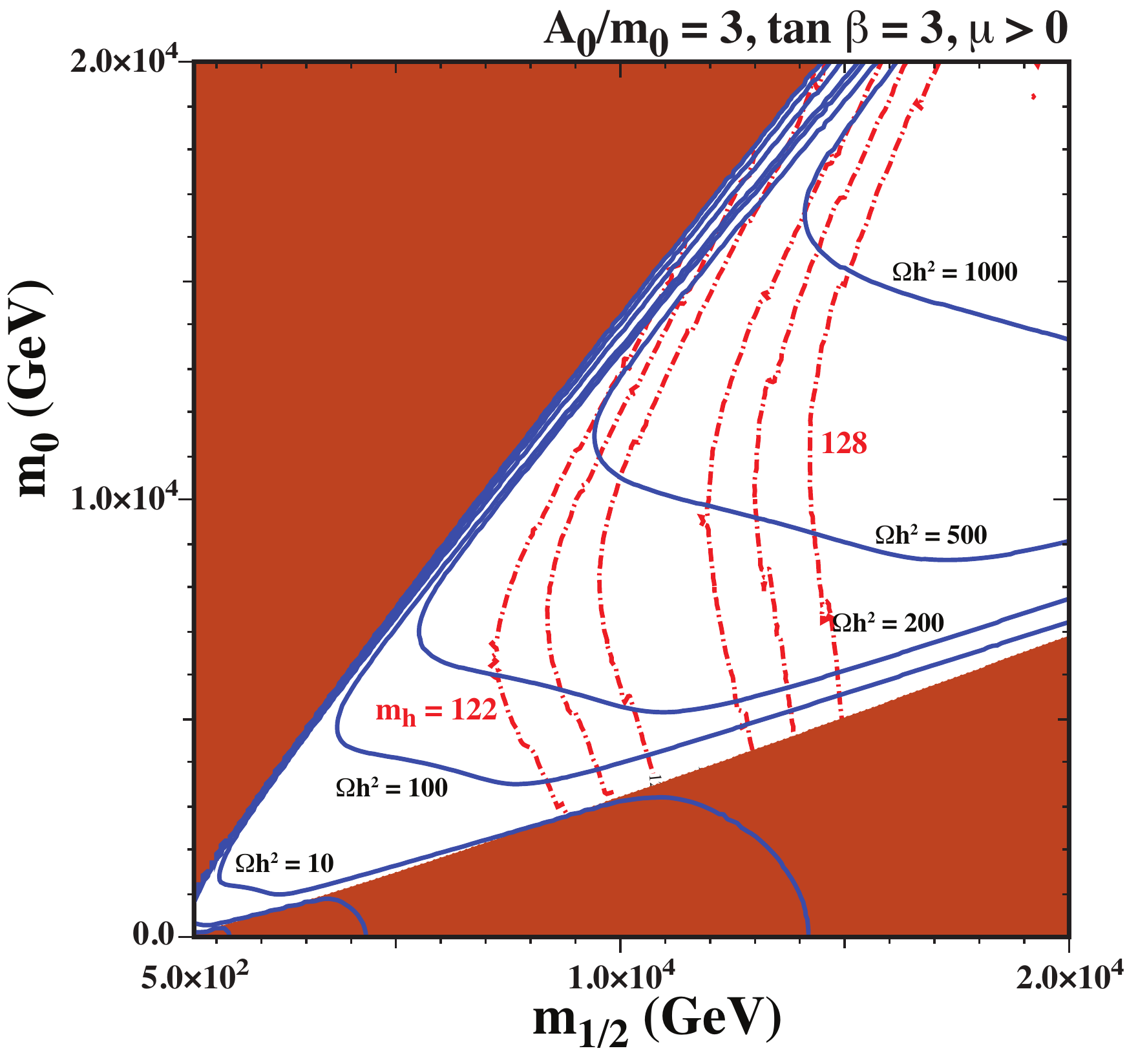}
%\hspace*{-0.7in}
\includegraphics[height=3in]{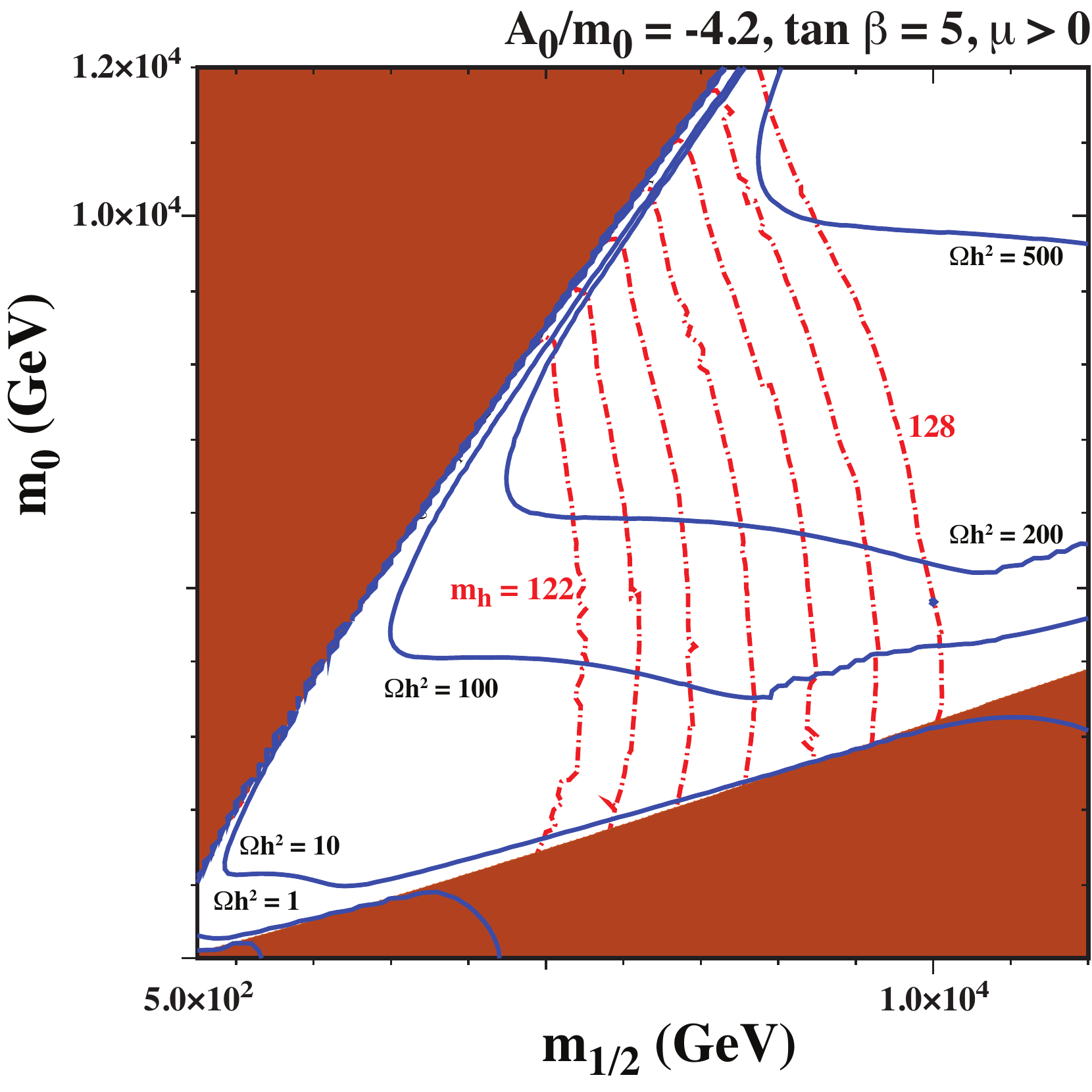}\\
\hfill\end{minipage}
\caption{
{\it
Some $(m_{1/2}, m_0)$ planes in standard SU(5) with $M_{in} = M_{GUT}$, $\tan \beta = 3$, $\mu > 0$, $A_0 = 0$,
(upper left panel), 
$M_{in} = M_{GUT}$, $\tan \beta = 10$, $\mu > 0$, $A_0 = 0$, (upper right panel),
$M_{in} = M_{GUT}$, $\tan \beta = 3$, $\mu > 0$, $A_0/m_0 = 3$, (lower left panel),$M_{in} = M_{GUT}$~GeV, $\tan \beta = 5$, $\mu > 0$, $A_0/m_0 = -4.2$, (lower right panel).
Here and in the subsequent Figure, the LSP (stau or stop) is charged in the dark red shaded regions, which are therefore excluded, there is no consistent electroweak vacuum in the pink shaded regions, the red dot-dashed lines are contours of $m_h$ calculated using {\tt FeynHiggs}~\protect\cite{FeynHiggs}, and the solid blue lines are contours of $\Omega_\chi h^2$ in the absence of subsequent entropy generation.
}}
\label{fig:CMSSM}
\end{figure}

Following this general discussion, we now consider some concrete examples
that illustrate the behaviour of the relic 
density as a function of the supersymmetric mass scales. We
first exhibit in Fig. \ref{fig:CMSSM} four examples of parameter planes in the CMSSM. 
In all four planes, the input universality scale is set to the GUT scale, taken to be the renormalization scale where the electroweak gauge couplings are equal, $g_1 = g_2$, and we assume that the Higgs mixing parameter $\mu > 0$. In the upper left panel, we choose $A_0/m_0 = 0$ and $\tan \beta = 3$. The dark red shaded region where $m_{1/2} \gg m_0$ is excluded because the lighter stau
is the LSP. The red dot-dashed lines are contours of constant Higgs masses between $m_h = 122$ and 128 GeV as calculated using {\tt FeynHiggs}~\cite{FeynHiggs}, which are consistent with the measured value within the calculational uncertainties. The solid blue contours
show values of the LSP relic density {\it in the absence of subsequent entropy generation}, $\Omega_\chi h^2$.
For this choice of $A_0$ and $\tan \beta$, in the absence of subsequent entropy generation the only viable area with $\Omega_\chi h^2 = 0.12$ would be a narrow strip at $m_{1/2} \lesssim 1$ TeV (outside the scale of this Figure)
lying very close to the line of mass degeneracy between the neutralino (a bino in this case) and the stau. 
Everywhere else in the plane $\Omega_\chi h^2$ is large
and varies between about 10 and 2000 for the parameter range shown. For an entropy release corresponding to $\Delta = 10^4$, the preferred region shifts to $m_{1/2} \approx 12$ TeV and $m_0 \approx 15$ TeV where $\Omega h^2 \approx 1000$, with $m_h = 125$ GeV. 

In the upper right panel, we show an analogous plane for larger $\tan \beta = 10$.  In this case, the pink shaded region at large $m_0 \gg m_{1/2}$ is excluded by the absence of a consistent electroweak vacuum. Just below this region, there is a focus-point strip with $\Omega_\chi h^2 = 0.1$, and for $m_{1/2} < 1$ TeV there is principle also
a narrow strip with the right relic density just above the stau LSP region. In the
conventional thermal freeze-out picture, values of $m_0(m_{1/2})$ are constrained to lie along one of these strips. 
However, the bulk of the plane has a much larger relic density,
which can reach $\sim 200$ when $m_h = 125$ GeV.
For this value of $\tan \beta$, $\Omega_\chi h^2 \sim 1000$
is found at very large $m_{1/2}$ where the Higgs mass is too large~\footnote{The position of the focus-point strip drifts up to higher $m_0$ for smaller $\tan \beta$, which is why it is not seen in panel a).}.

The lower two panels show similar patterns
when $A_0/m_0 = 3$ with $\tan \beta = 3$ (lower left)
and $A_0/m_0 = -4.2$ with $\tan \beta = 5$ (lower right).
In both panels, in addition to the stau LSP region (lower
wedges), there are also stop LSP regions at large $m_0/m_{1/2}$. 
Adjacent to each of the stop LSP regions, there is a stop
coannihilation strip with $\Omega_\chi h^2 = 0.12$, which
is invisibly thin, because of the small uncertainty in the Planck 
determination of $\Omega_\chi h^2$ and the scale of the Figure. 
Once again, we see that the introduction of entropy
opens up the plane so that one is no longer confined to 
these narrow coannihilation strips~\footnote{As previously, there is also an invisible stau coannihilation strip at small $m_{1/2}$.}.

%%%%%
%%%%%%%%%%%%%%%%%%%%%%%%%%%%%%%%%%%%%%%%%%%%%%%%%%%%%

\begin{figure}[htb!]
%\vskip 0.5in
%\vspace*{-0.75in}
%\hspace*{-.70in}
\begin{minipage}{8in}
%\hspace*{-1.2in}
\includegraphics[height=3in]{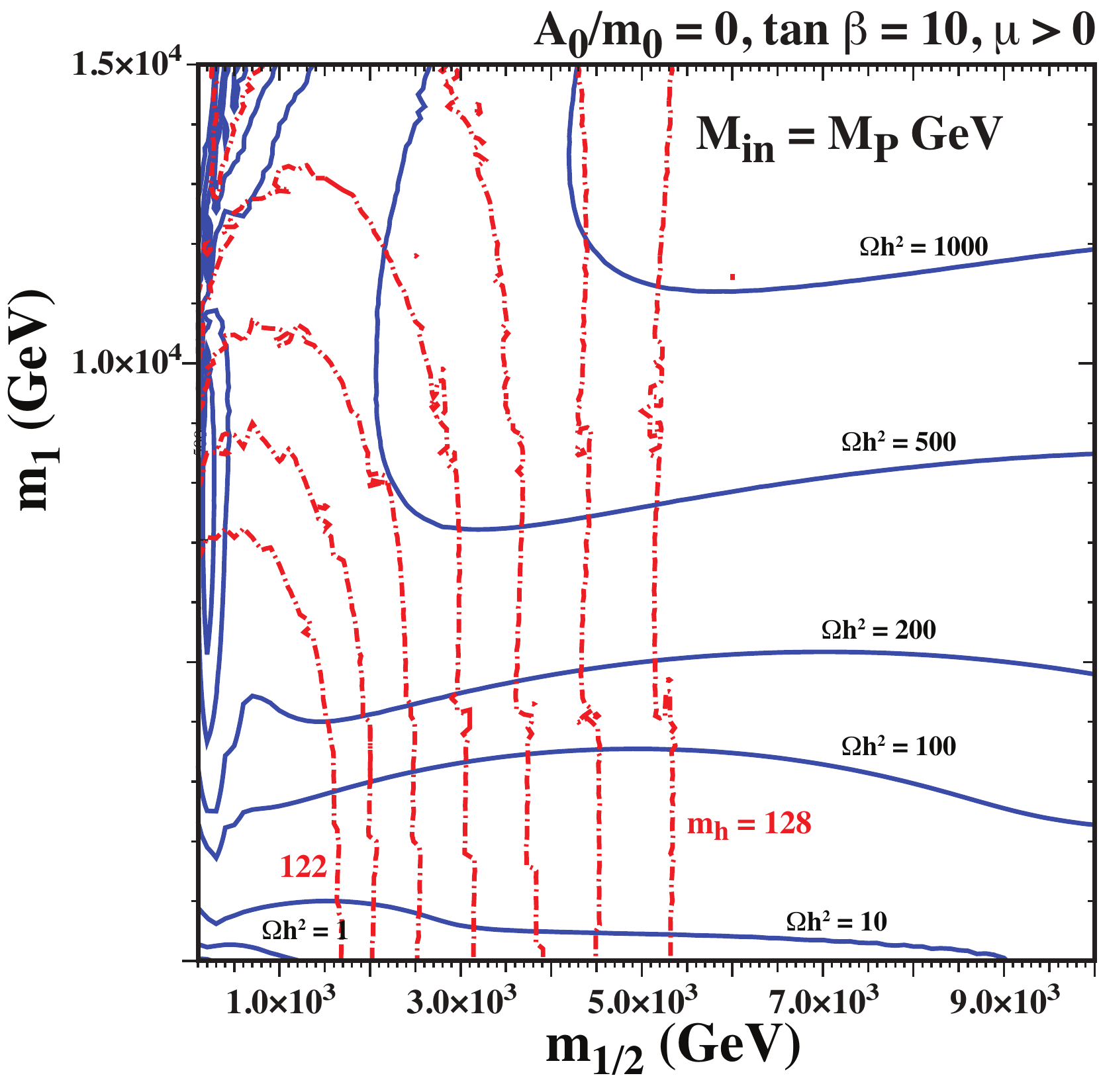}
%\hspace*{-2.7in}
\includegraphics[height=3in]{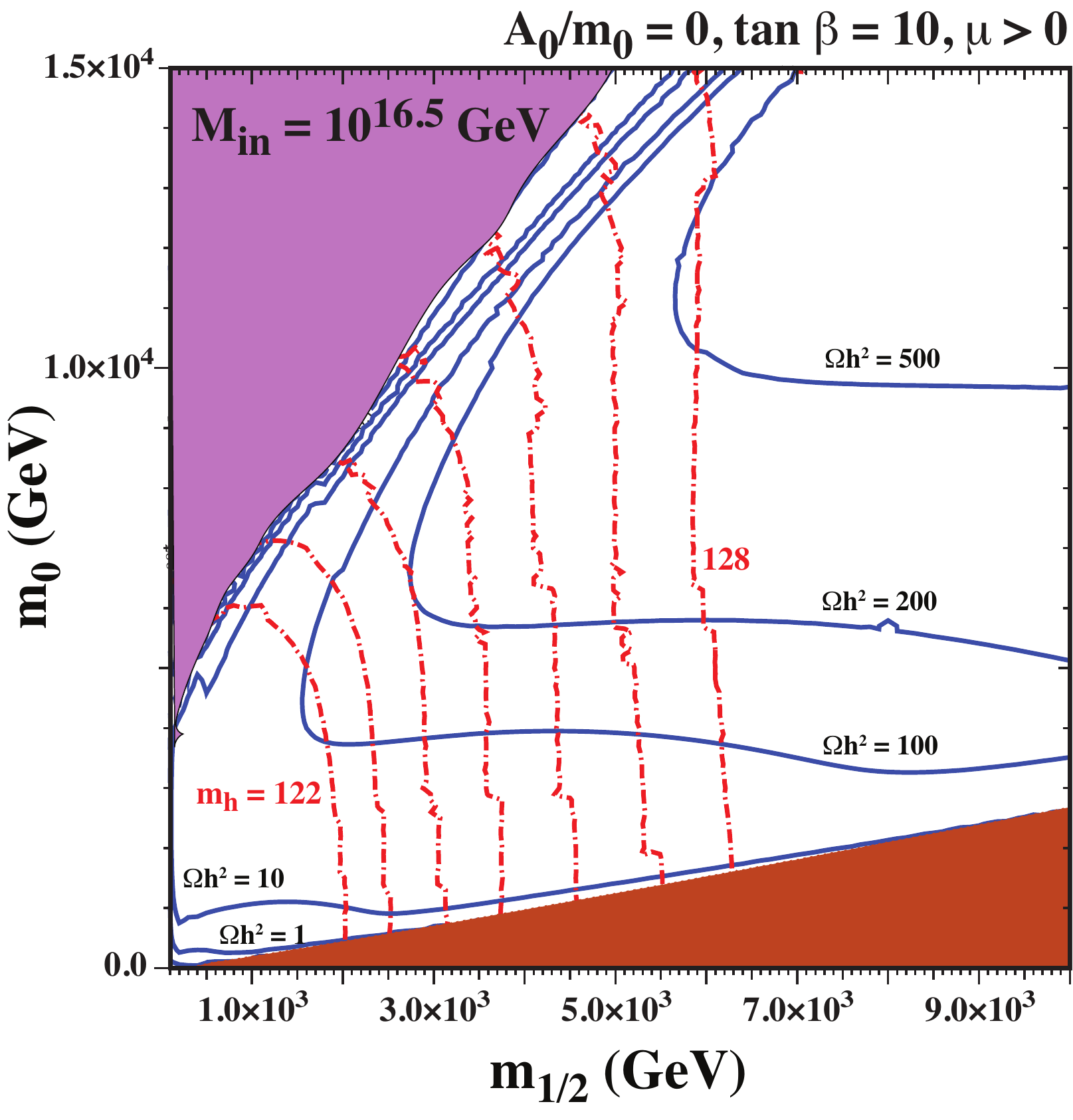}\\
%\vspace*{-1in}
\hfill
%\vspace*{-2in}
\includegraphics[height=3in]{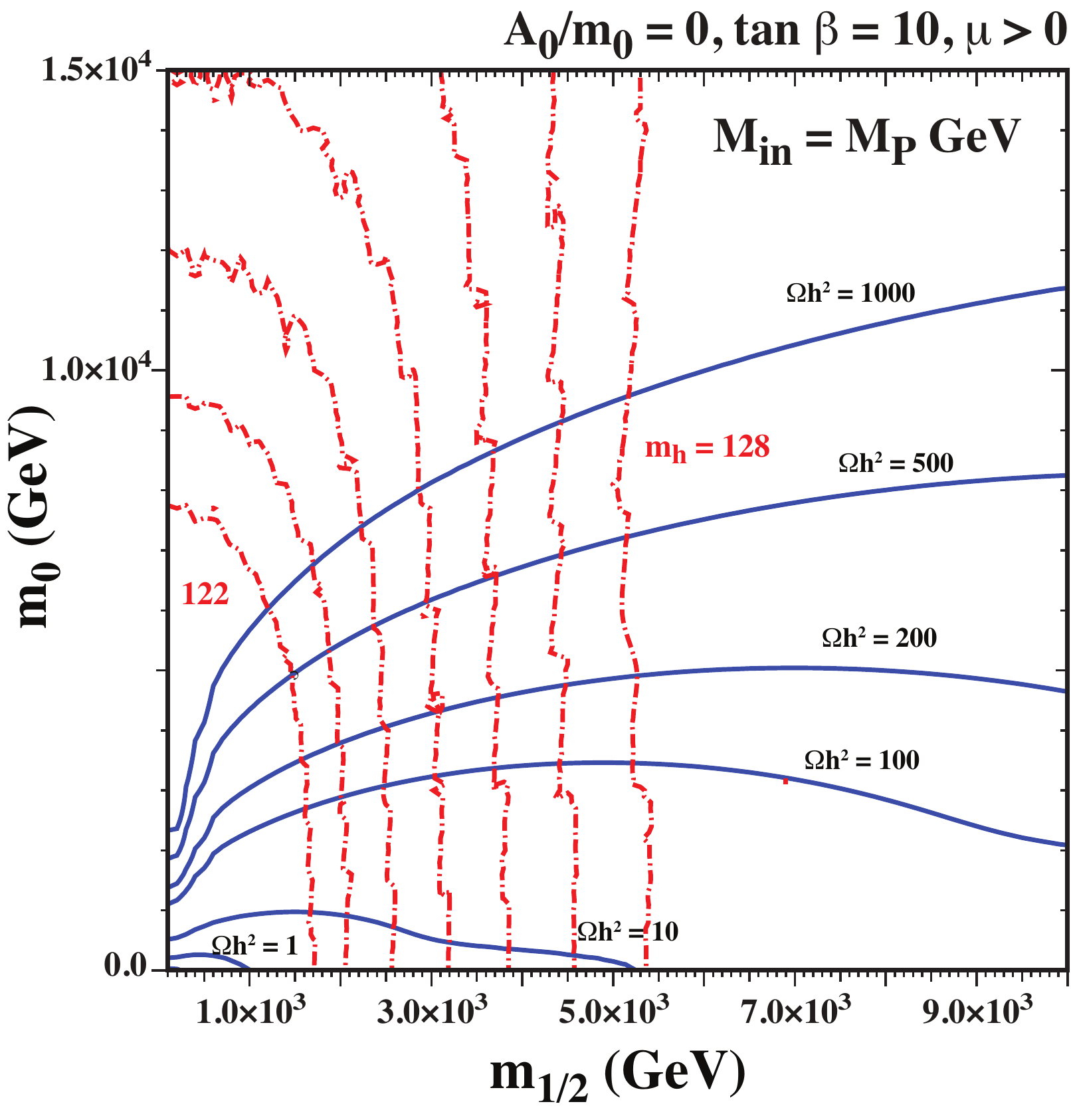}
%\hspace*{-0.7in}
\includegraphics[height=3in]{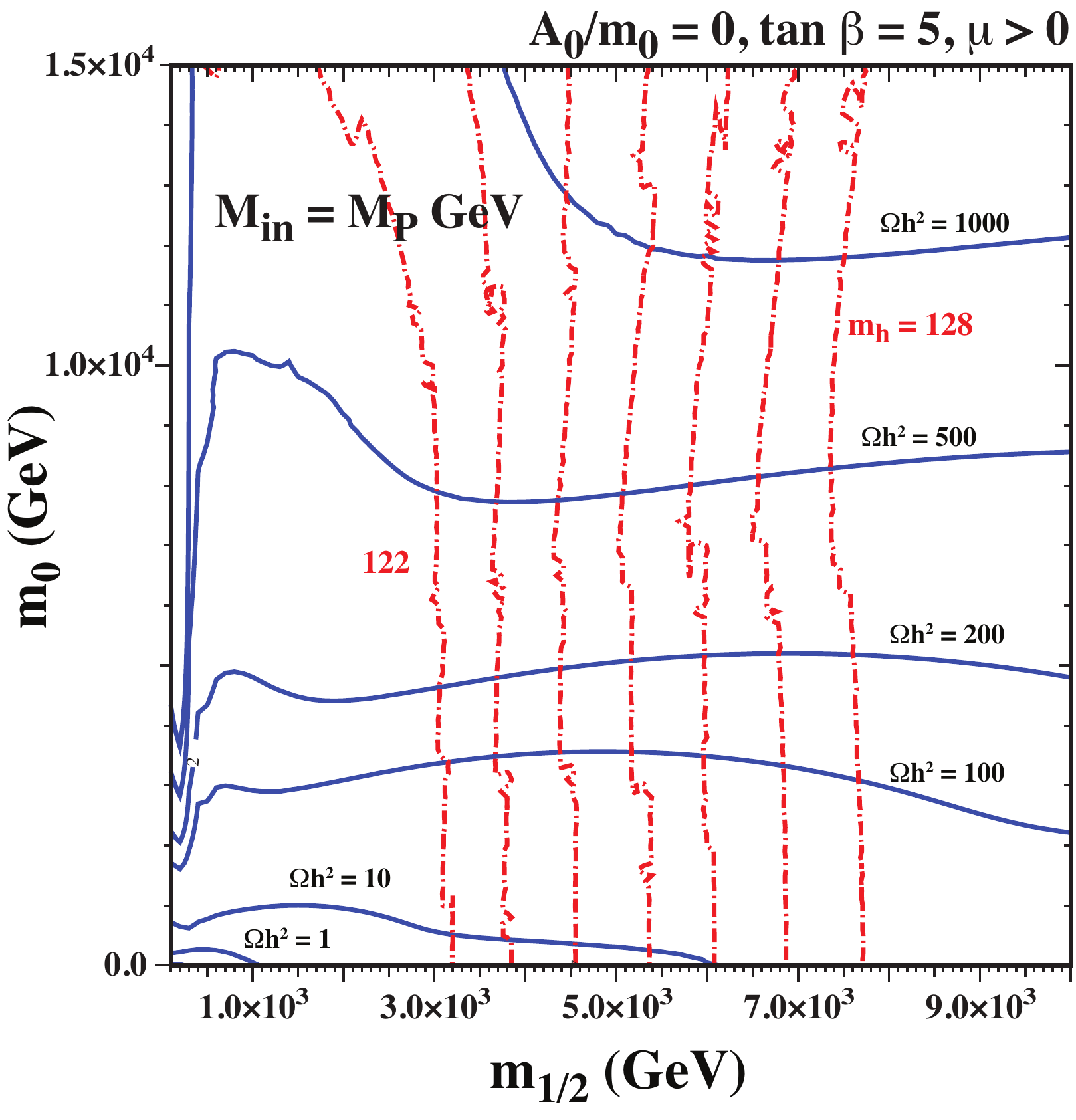}
\hfill\end{minipage}
\caption{
{\it
Some $(m_{1/2}, m_0)$ planes in the flipped super-GUT model with $M_{in} = M_P$, $\tan \beta = 10$, $\mu > 0$, $A_0 = 0$, {\boldmath${ \lambda}$} $=(0.3, 0.1)$ (upper left panel), $M_{in} = 10^{16.5}$~GeV, $\tan \beta = 10$, $\mu > 0$, $A_0 = 0$, {\boldmath${ \lambda}$} $= (0.3, 0.1)$ (upper right panel),
$M_{in} = M_P$, $\tan \beta = 10$, $\mu > 0$, $A_0 = 0$, {\boldmath${ \lambda}$} $= (0.3, 0.3)$ (lower left panel),
$M_{in} = M_P$, $\tan \beta = 5$, $\mu > 0$, $A_0 = 0$, {\boldmath${ \lambda}$} $= (0.3, 0.1)$ (lower right panel). The shadings and line styles are the same as in the previous Figure. These figures are not sensitive to the
choice of $\lambda_6$.
}}
\label{fig:superGUT1}
\end{figure}

The above four planes are rather generic for the CMSSM, and there are similar features in the 
super-GUT CMSSM based on flipped SU(5) as originally considered in \cite{emo3}. In all four flipped super-GUT planes shown in Fig.~\ref{fig:superGUT1}, we take
$A_0/m_0 = 0$ and $\mu > 0$. The upper left panel is 
similar to one considered in \cite{emo3}, and the input
universality scale is chosen to be the Planck scale
with $\tan \beta = 10$. The flipped SU(5) couplings 
are chosen as {\boldmath${ \lambda}$} $= (\lambda_4, \lambda_5) = (0.3,0.1)$. These planes are not sensitive to the choice of $\lambda_6$. There is 
a region with $\Omega_\chi h^2 = 0.12$ due to a rapid
$s$-channel Higgs annihilation funnel at very low $m_{1/2} \sim 200$ GeV, which is responsible for the structures in the relic
density contours at low $m_{1/2}$.  However, in the bulk of the plane, as in the CMSSM, the relic density is significantly higher, particularly when $m_h = 125$ GeV. In the upper
right panel of Fig.~\ref{fig:superGUT1}, we have taken the input universality scale
at $M_{in} = 10^{16.5}$ GeV, and the resulting plane similar to the case shown in the upper right panel of Fig. \ref{fig:CMSSM}.  

In the lower left panel of Fig. \ref{fig:superGUT1}, 
we have increased $\lambda_5$ so that {\boldmath${ \lambda}$} $=  (0.3,0.3)$ with the same choices of the other parameter as used in the upper left panel. In this case,
$\Omega_\chi h^2$ easily reaches $\mathcal{O}$(1000) when
$m_h = 125$ GeV. Similarly, we show in the lower right panel
the plane with $\tan \beta = 5$ with {\boldmath${ \lambda}$} $=  (0.3,0.1)$,
where values of $\Omega_\chi h^2 \gtrsim 1000$ are again attained.

In summary, when the supersymmetric soft mass scales 
are taken generically to be of order 10 TeV
as illustrated in the examples discussed above, 
the resulting relic density from thermal freeze-out
is $\gg 1$ in the absence of subsequent entropy generation, and easily reaches $\mathcal{O}$(1000) when 
$m_h = 125$ GeV. However, in the presence of a factor ${\cal O}(10^4)$ of
entropy generation, as advocated in the previous Section, the correct cold dark matter
relic density can be obtained in generic domains of the parameter space of our model.  If the contribution to $\Omega_\chi h^2$
from gravitino decay is significant, the thermal
component would necessarily have to be smaller,
but there is no apparent need for fine-tuning.

%%%%%
%%%%%%%%%%%%%%%%%%%%%%%%%%%%%%%%%%%%%%%%%%%%%%%%%%%%%

%%%%%%%%%%%%%%%%%%%%%%%
\section{Conclusions}
\label{sec:conclusion}
%%%%%%%%%%%%%%%%%%%%%%%

In this paper we have developed further a string-inspired model for particle cosmology proposed previously~\cite{egnno4}, based on a flipped SU(5)$\times$U(1) gauge group embedded in no-scale supergravity, as outlined in Fig.~\ref{fig:concept}. We have paid particular attention to incorporating all the relevant cosmological constraints, including the realization of Starobinsky-like inflation, a baryon density $n_B/s$ and a cold dark matter density $\Omega_{\rm CDM} h^2$ consistent with observations, light neutrino masses and mixing parameters consistent with data on astrophysical structures as well as oscillation measurements, and the successful realization of Big Bang nucleosynthesis. 

It is a striking feature of the model that the generation of an entropy factor $\Delta \sim 10^4$ is not only required for the consistency of $n_B/s$ and $\Omega_{\rm CDM} h^2$ with observations, but is also to be expected during the flaton phase transition in the early universe. Within our model framework, this large increase in entropy requires relatively heavy supersymmetric particles weighing ${\cal O}(10)$~TeV. In this case the dark matter density calculated via the conventional freeze-out mechanism in conventional cosmology would, in the absence of fine-tuning, be orders of magnitude greater than the small range allowed by Planck and other measurements. However, the large entropy factor reduces the cold dark matter density into the Planck range, maintaining consistency with relatively heavy supersymmetric particles weighing ${\cal O}(10)$~TeV. Therefore, within our model framework it is no surprise that the LHC has not (yet) discovered any supersymmetric particles. However, they may well lie within reach of a next-generation ${\cal O}(100)$~TeV proton-proton collider such as FCC-hh~\cite{FCC-hh}.

There are many important aspects of our model that remain to be worked out. For example, whilst our model yields a small value of the tensor-to-scalar perturbation ratio $r$ in the cosmic microwave background that is highly consistent with present observations, and the model prediction for the value of the scalar tilt, $n_s$, is also consistent with Planck and other data at the 68\% CL~\cite{Aghanim:2018eyx}. Another area where the model could make interesting predictions worthy of a dedicated study is that of baryon decay. Also, it would be interesting to make a global fit to the parameters of the model in an effort to pin them down more narrowly. Taking a broader perspective, it would also be interesting to explore the generality of some of the features we have found in this model. For example, how general is the expectation of substantial entropy generation, and the consequent opening of the range of generic sparticle masses to ${\cal O}(10)$ TeV? We shall certainly be returning to some of these issues in future work.

%%%%%%%%%%%%%%%%%%%%%%%%%%%%%
\section*{Acknowledgements}
%%%%%%%%%%%%%%%%%%%%%%%%%%%%%
The work of J.E.~was supported partly by the United Kingdom STFC Grant ST/P000258/1 
and partly by the Estonian Research Council via a Mobilitas Pluss grant. The work of M.A.G.G.~ was supported by the Spanish Agencia Estatal de Investigaci\'on through the grants FPA2015-65929-P (MINECO/FEDER, UE), PGC2018095161-B-I00, IFT Centro de Excelencia Severo Ochoa SEV-2016-0597, and Red Consolider MultiDark FPA2017-90566-REDC. The work of N.N.~was supported by the Grant-in-Aid for Young Scientists B (No.17K14270) and Innovative Areas (No.18H05542). The work of D.V.N.~was supported partly by the DOE grant DE-FG02-13ER42020 
and partly by the Alexander S. Onassis Public Benefit Foundation. The work of K.A.O.~was supported partly
by the DOE grant DE-SC0011842 at the University of Minnesota and
 acknowledges support by the Director, Office of Science, Office of High Energy Physics of the U.S. Department of Energy under the Contract No. DE-AC02-05CH11231.
K.A.O. would also like to thank the Department of Physics and the 
high energy theory group
at the University of California, Berkeley as well as the theory group at LBNL
for their hospitality and financial support while
finishing this work.

%%%%%%%%%%%%%%%%%%%%%%%%%%%%%

\end{document}